\documentclass[usenatbib,twocolumn]{aastex7}

\usepackage{natbib}
\shorttitle{Dust in SN 1995N}
\shortauthors{Clayton et al.}

\begin{document}

\title{Very Late-Time JWST and Keck Spectra of the Oxygen-Rich Supernova 1995N}



\author[0000-0002-0141-7436]{Geoffrey C. Clayton}
\affiliation{Space Science Institute,
4765 Walnut St., Suite B
Boulder, CO 80301, USA}
\email[show]{
gclayton@spacescience.org}
\affiliation{Department of Physics \& Astronomy, Louisiana State University, Baton Rouge, LA 70803, USA}
\affiliation{Maria Mitchell Association, 4 Vestal St., Nantucket, MA 02554, USA}

\author[0000-0002-4000-4394]{R. Wesson}
\affiliation{School of Physics and Astronomy, Cardiff University, Queens Buildings, The Parade, Cardiff CF24 3AA, UK}
\email{rw@nebulousresearch.org}

\author[0000-0003-2238-1572]{Ori D. Fox}
\affiliation{Space Telescope Science Institute, 3700 San Martin Drive, Baltimore, MD 21218, USA}
\email{ofox@stsci.edu}

\author[0000-0002-9301-5302]{Melissa Shahbandeh}
\affiliation{Space Telescope Science Institute, 3700 San Martin Drive, Baltimore, MD 21218, USA}
\affiliation{Department of Physics and Astronomy, Johns Hopkins University, 3400 North Charles Street, Baltimore, MD 21218, USA}
\email{mshahbandeh@stsci.edu}

\author[0000-0003-3460-0103]{Alexei V. Filippenko}
\affiliation{Department of Astronomy, University of California, Berkeley, CA 94720-3411, USA}
\email{afilippenko@berkeley.edu}

 \author[0000-0002-9915-1372]{Bryony Nickson}
 \affiliation{Space Telescope Science Institute, 3700 San Martin Drive, Baltimore, MD 21218, USA}
 \email{bnickson@stsci.edu}

 \author[0000-0003-0209-674X]{Michael Engesser}
 \affiliation{Space Telescope Science Institute, 3700 San Martin Drive, Baltimore, MD 21218, USA}
 \email{mengesser@stsci.edu}

\author[0000-0001-9038-9950]{Schuyler D.~Van Dyk}
\affiliation{Caltech/IPAC, Mailcode 100-22, Pasadena, CA 91125, USA}
\email{vandyk@ipac.caltech.edu}

\author[0000-0002-2636-6508]{WeiKang Zheng}
\affiliation{Department of Astronomy, University of California, Berkeley, CA 94720-3411, USA}
\email{weikang@berkeley.edu}

\author[0000-0001-5955-2502]{Thomas G. Brink}
\affiliation{Department of Astronomy, University of California, Berkeley, CA 94720-3411, USA}
\email{tgbrink@berkeley.edu}

\author[0000-0002-6535-8500]{Yi Yang}
\affiliation{Physics Department, Tsinghua University, Beijing 100084, People’s Republic of China}
\email{yiyangtamu@gmail.com}

\author[0000-0001-7380-3144]{Tea Temim}
\affiliation{Department of Astrophysical Sciences, Princeton University, Princeton, NJ 08544, USA}
\email{temim@astro.princeton.edu}

\author[0000-0001-5510-2424]{Nathan Smith}
\affiliation{Steward Observatory, University of Arizona, 933 N. Cherry St, Tucson, AZ 85721, USA}
\email{nathans@as.arizona.edu}

\author[]{Jennifer Andrews}
\affiliation{Gemini Observatory, 670 N. Aohoku Place, Hilo, HI 96720,
USA}
\email{Jennifer.Andrews@noirlab.edu}

\author[0000-0002-5221-7557]{Chris Ashall}
\affiliation{Institute for Astronomy, University of Hawai’i at Manoa, 2680 Woodlawn Dr., Honolulu, HI 96822-1839, USA}
\email{cashall@hawaii.edu}

\author[]{Ilse De Looze}
\affiliation{Sterrenkundig Observatorium, Ghent University, Krijgslaan 281 – S9, 9000 Gent, Belgium}
\email{Ilse.DeLooze@UGent.be}

\author[0000-0002-7566-6080]{James M. Derkacy}
\affiliation{Space Telescope Science Institute, 3700 San Martin Drive, Baltimore, MD 21218, USA}
\email{jmderkacy@gmail.com}

\author[]{Luc Dessart}
\affiliation{Institut d’Astrophysique de Paris, CNRS–Sorbonne Université, 98
bis boulevard Arago, F-75014 Paris, France}
\email{dessart@iap.fr}

\author[]{Michael Dulude}
\affiliation{Space Telescope Science Institute, 3700 San Martin Drive, Baltimore, MD 21218, USA}
\email{dulude@stsci.edu}

\author[]{Eli Dwek}
\affiliation{Observational Cosmology Lab, NASA Goddard Space Flight Center, Code 665, Greenbelt, MD 20771, USA}
\email{eli.dwek@nasa.gov}

\author[0000-0002-2445-5275]{Ryan J. Foley}
\affiliation{Department of Astronomy and Astrophysics, University of California, Santa Cruz, CA 95064, USA}
\email{foley@ucsc.edu}

\author[]{Suvi Gezari}
\affiliation{Space Telescope Science Institute, 3700 San Martin Drive, Baltimore, MD 21218, USA}
\email{sgezari@stsci.edu}

\author[0000-0001-6395-6702]{Sebastian Gomez}
\affiliation{Space Telescope Science Institute, 3700 San Martin Drive, Baltimore, MD 21218, USA}
\email{sgomez@stsci.edu}

\author[]{Shireen Gonzaga}
\affiliation{Space Telescope Science Institute, 3700 San Martin Drive, Baltimore, MD 21218, USA}
\email{shireen@stsci.edu}

\author[]{Siva Indukuri}
\affiliation{Department of Physics and Astronomy, Johns Hopkins University, 3400 North Charles Street, Baltimore, MD 21218, USA}
\email{sinduku1@jh.edu}

\author[0000-0001-5754-4007]{Jacob Jencson}
\affiliation{IPAC, Mail Code 100-22, Caltech, 1200 E. California Blvd., Pasadena, CA 91125}
\email{jjencson@ipac.caltech.edu}

\author[0000-0001-5975-290X]{Joel Johansson}
\affiliation{Oskar Klein Centre, Department of Physics, Stockholm University,
AlbaNova, SE-10691 Stockholm, Sweden}
\email{joeljo@fysik.su.se}

\author[]{Mansi Kasliwal}
\affiliation{Cahill Center for Astrophysics, California Institute of Technology, 1200 E. California Blvd. Pasadena, CA 91125, USA}
\email{mansi@astro.caltech.edu}

\author[0009-0003-8380-4003]{Zachary G. Lane}
\affiliation{School of Physical and Chemical Sciences — Te Kura Matu, University of Canterbury, Private Bag 4800, Christchurch 8140, New Zealand}
\email{zachary.lane@pg.canterbury.ac.nz}

\author[]{Ryan Lau}
\affiliation{NSF’s NOIRLab, 950 N. Cherry Avenue, Tucson, AZ 85719, USA}
\email{ryan.lau@noirlab.edu}

\author[]{David Law}
\affiliation{Space Telescope Science Institute, 3700 San Martin Drive, Baltimore, MD 21218, USA}
\email{dlaw@stsci.edu}

\author[0000-0001-5788-5258]{Anthony Marston}
\affiliation{European Space Agency (ESA), ESAC, 28692 Villanueva de la Canada, Madrid, Spain}
\email{anmarston@stsci.edu}

\author[0000-0002-0763-3885]{Dan Milisavljevic}
\affiliation{Purdue University, Department of Physics and Astronomy, 525 Northwestern Ave, West Lafayette, IN 4790720, USA}
\email{dmilisav@purdue.edu}

\author[0000-0002-2432-8946]{Richard O'Steen}
\affiliation{Space Telescope Science Institute, 3700 San Martin Drive, Baltimore, MD 21218, USA}
\email{rosteen@stsci.edu}

\author[0000-0002-2361-7201]{Justin Pierel}
\affiliation{Space Telescope Science Institute, 3700 San Martin Drive, Baltimore, MD 21218, USA}
\email{jpierel@stsci.edu}

\author[0000-0002-4410-5387]{Armin Rest}
\affiliation{Department of Physics and Astronomy, Johns Hopkins University, 3400 North Charles Street,
Baltimore, MD 21218, USA}
\affiliation{Space Telescope Science Institute, 3700 San Martin Drive, 
Baltimore, MD 21218, USA}
\email{arest@stsci.edu}

\author[0000-0002-9820-679X]{Arkaprabha Sarangi}
\affiliation{DARK, Niels Bohr Institute, University of Copenhagen, Jagtvej 128,
2200 Copenhagen, Denmark}
\email{00000144d9e8395f-dmarc-request@MAILLIST.STSCI.EDU}

\author[0000-0003-2445-3891]{Matthew Siebert}
\affiliation{Space Telescope Science Institute, 3700 San Martin Drive, Baltimore, MD 21218, USA}
\email{msiebert@stsci.edu}

\author[]{Michael Skrutskie}
\affiliation{Department of Astronomy, University of Virginia, Charlottesville, VA 22904-4325, USA}
\email{mfs4n@virginia.edu}

\author[0000-0002-7756-4440]{Lou Strolger}
\affiliation{Space Telescope Science Institute, 3700 San Martin Drive, Baltimore, MD 21218, USA}
\email{strolger@stsci.edu}

\author[0000-0003-4610-1117]{Tam\'as Szalai}
\affiliation{Department of Experimental Physics, Institute of Physics, University of Szeged, D\'om t\'er 9, 6720 Szeged, Hungary}
\affiliation{MTA-ELTE Lendület ``Momentum" Milky Way Research Group,
Szent Imre H. st. 112, 9700 Szombathely, Hungary}
\email{szaszi@titan.physx.u-szeged.hu}

\author[0000-0002-1481-4676]{Samaporn Tinyanont}
\affiliation{National Astronomical Research Institute of Thailand (NARIT), Chiang Mai, 50180, Thailand}
\email{samaporn@narit.or.th}

\author[]{Qinan Wang}
\affiliation{Department of Physics and Kavli Institute for Astrophysics and Space Research, Massachusetts Institute of Technology, 77 Massachusetts Avenue, Cambridge, MA 02139, USA}
\email{qwang75@jhu.edu}

\author[]{Brian Williams}
\affiliation{Observational Cosmology Lab, NASA Goddard Space Flight Center, Code 665, Greenbelt, MD 20771, USA}
\email{brian.j.williams@nasa.gov}

\author[]{Lin Xiao}
\affiliation{Department of Physics, College of Physical Sciences and Technology, Hebei University, Baoding 071002, China}
\email{linxiao@hbu.edu.cn}

\author[0000-0001-7473-4208]{Szanna Zs{\'i}ros}
\affiliation{Department of Experimental Physics, Institute of Physics, University of Szeged, D\'om t\'er 9, 6720 Szeged, Hungary}
\affiliation{HUN-REN CSFK Konkoly Observatory, Konkoly Thege M. ut 15-17, Budapest, 1121, Hungary}
\email{szannazsiros@titan.physx.u-szeged.hu}

\begin{abstract}
We present new {\it JWST}/MIRI MRS and Keck spectra of SN 1995N obtained in 2022--2023, more than 10,000 days after the supernova (SN) explosion. These spectra are among the latest direct detections of a core-collapse SN, both through emission lines in the optical and thermal continuum from infrared dust emission. The new infrared data show that dust heating from radiation produced by the ejecta interacting with circumstellar matter is still present, but greatly reduced from when SN 1995N was observed by the {\it Spitzer Space Telescope} and {\it WISE} in 2009/2010 and 2018, when the dust mass was estimated to be 0.4\,M$_{\sun}$. New radiative-transfer modeling suggests that the dust mass and grain size may have increased between 2010 and 2023. The new data can alternatively be well fit with a dust mass of 0.4\,M$_{\sun}$ and a much reduced heating source luminosity. The new late-time spectra show unusually strong oxygen forbidden lines, stronger than the H$\alpha$ emission. 
This indicates that SN 1995N may have exploded as a stripped-envelope SN which then interacted with a massive H-rich
circumstellar shell, changing it from intrinsically Type Ib/c to Type IIn. 
The late-time spectrum results when the reverse shock begins to excite the inner H-poor, O-rich ejecta. This change in the spectrum is rarely seen, but marks the start of the transition from SN to SN remnant.

\end{abstract}


\section{Introduction}
Core-collapse supernovae (CCSNe) are likely a major contributor of dust in the high-redshift universe. 
Submillimeter observations indicate that early galaxies may contain up to 10$^8$\,M$_{\sun}$ of
dust at redshift $z \geq 6$  \citep[e.g.,][]{Bertoldi_2003,Dwek_2014,2025MNRAS.tmp..117C}.
If true, the dust budget requires $\sim$1\,M$_{\odot}$ of dust to form in each SN \citep{Dwek_2007}.  

In the last two decades, many nearby CCSNe have been studied to determine how much dust these objects can produce. Most studies of the mass of dust associated with CCSNe have been 2--3 orders of magnitude too small \citep[e.g.,][and references therein]{zsiros24}. However, more recent observations enabled by larger telescopes and new technologies are allowing us to re-examine this perspective. For example, mid-infrared (IR) observations revealed at least 0.3\,M$_{\sun}$ of dust in supernova remnant (SNR) G54.1+0.3 \citep{Temim_2017}, while observations in the far-IR with the {\it Herschel Space Observatory} found 0.4--0.7\,M$_{\sun}$ of cold dust ($\sim$20\,K) in the ejecta of SN 1987A \citep{Matsuura_11,Matsuura_15}, 0.1--0.6\,M$_{\sun}$ in the Cas~A SNR \citep{Barlow_2010,De_Looze_2017}, and 0.02--0.4\,M$_{\sun}$ in the Crab Nebula SNR \citep{Gomez_12, Temim_2013, Owen_15}. More recently, {\it James Webb Space Telescope (JWST)} Mid-Infrared Instrument (MIRI) imaging of the Type IIP SN~2004et and Medium Resolution Spectroscopy (MRS) of the Type IIn SN~2005ip uncovered one of the largest newly formed dust masses in an extragalactic SN besides SN 1987A \citep{shahbandeh23,2024arXiv241009142S}.

There are two competing scenarios for how these large masses of dust might be formed. The first suggests that there is continuous dust formation in the ejecta so the dust mass can continue to increase for decades \citep{2014Natur.511..326G, 2015MNRAS.446.2089W}, while the second posits that large amounts of dust form at early times hidden in dense clumps \citep{2015ApJ...810...75D,2019ApJ...871L..33D}. The results for SN 2005ip also raise the possibility that the post-shock dust environment in some interacting SNe might be most conducive to dust formation \citep{2024arXiv241009142S}. Therefore, extremely late-time measurements of CCSN dust are valuable, but also quite rare given the faint nature of SNe at late times. 

SN 1995N was discovered on 1995 May 5 (UTC dates are used throughout this paper) in the galaxy MCG-02-38-17 (Arp 261), and spectra identified it as a peculiar Type II SN  \citep{1995IAUC.6170....1P,1995IAUC.6174....1G},
showing relative narrow emission components characteristic of SNe~IIn  \citep{2002ApJ...572..350F}.
%
%
It was bright at X-ray and radio wavelengths at early times \citep{1996IAUC.6386....1V,1996IAUC.6445....1L}, 
making it a member of a small subset of SNe IIn with SNe 1978K, 1986J,  1988Z, and 1998S \citep{2002ApJ...572..350F}.
SN 1995N is more radio luminous than many of the observed SNe IIn
 but less luminous than SN 1988Z \citep{2009ApJ...690.1839C}.
Its high luminosity and flux evolution imply interaction of the ejecta with inhomogeneous circumstellar matter \citep[CSM;][]{2000MNRAS.319.1154F}. 
The CSM is a gas and dust shell lost by the star before the SN explosion.
X-ray observations obtained $\sim$9\,yr after the explosion show that the flux dropped by an order of magnitude in 6\,yr and could be well fit by thermal emission from CSM interaction, and the higher spatial resolution of the {\it Chandra X-ray Observatory}, shows that the X-ray light curve of SN 1995N is consistent with a linear decline \citep{2005MNRAS.364.1419Z,2005ApJ...629..933C}.
SN 1995N was monitored by the VLA for 11\,yr beginning soon after the explosion. The radio emission, much like the X-rays, is consistent with bremsstrahlung radiation  \citep{2009ApJ...690.1839C}.


In this paper, we report new {\it JWST} and Keck spectra of SN 1995N obtained over 10,000\,d after the SN explosion.
SN 1995N is an ideal target for late-time observations, being on the outskirts of its host galaxy
and having an extremely slow decline in luminosity \citep{2011AN....332..266P}.

\begin{figure*}
\centering
\includegraphics[width=6in]{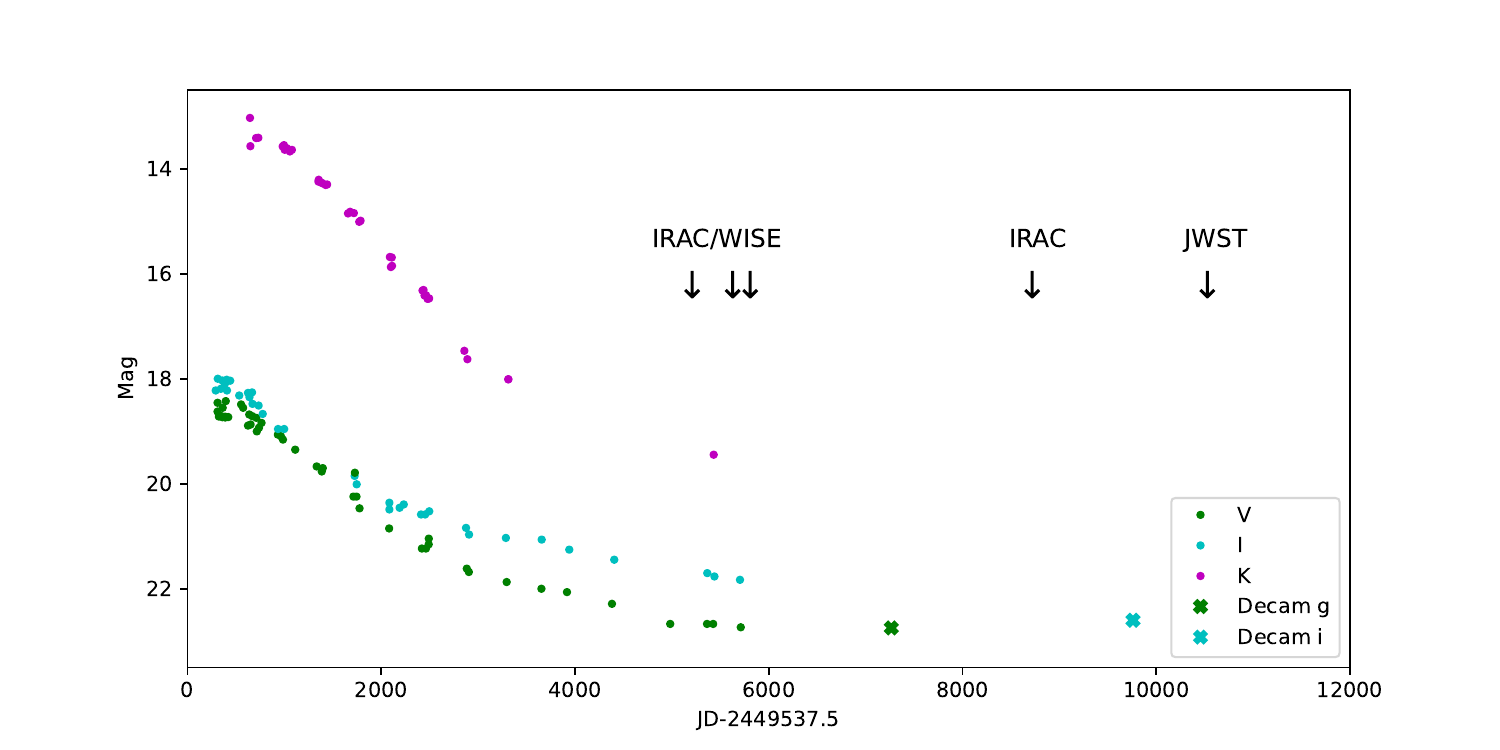}
\caption{Light curve of SN 1995N covering 26\,yr after the explosion \citep{2005ASPC..342..285P,2011AN....332..266P}. The arrows show the epochs of the IR observations with {\it Spitzer}/IRAC, {\it WISE},  and {\it JWST}.}
\label{fig:lightcurve}
\end{figure*}
\section{Existing Observations}
\subsection{Photometry}

Almost 27\,yr of monitoring of SN 1995N in the $UBVRIJHK$ bands is shown in Figure~\ref{fig:lightcurve} \citep{1995IAUC.6170....1P,1999IAUC.7141....3S,2001IAUC.7626....3S,Gerardy_2002,2002PASP..114..403L,2005ASPC..342..285P,2005MNRAS.364.1419Z}. Photometry from 2MASS, Pan-STARRS, VLT-ATLAS, and DENIS is also plotted.
The object faded by only about $\Delta V = 1.2$\,mag in the first 4\,yr after the explosion  \citep{2002PASP..114..403L}.

\begin{figure*}
\centering
\includegraphics[width=5in]{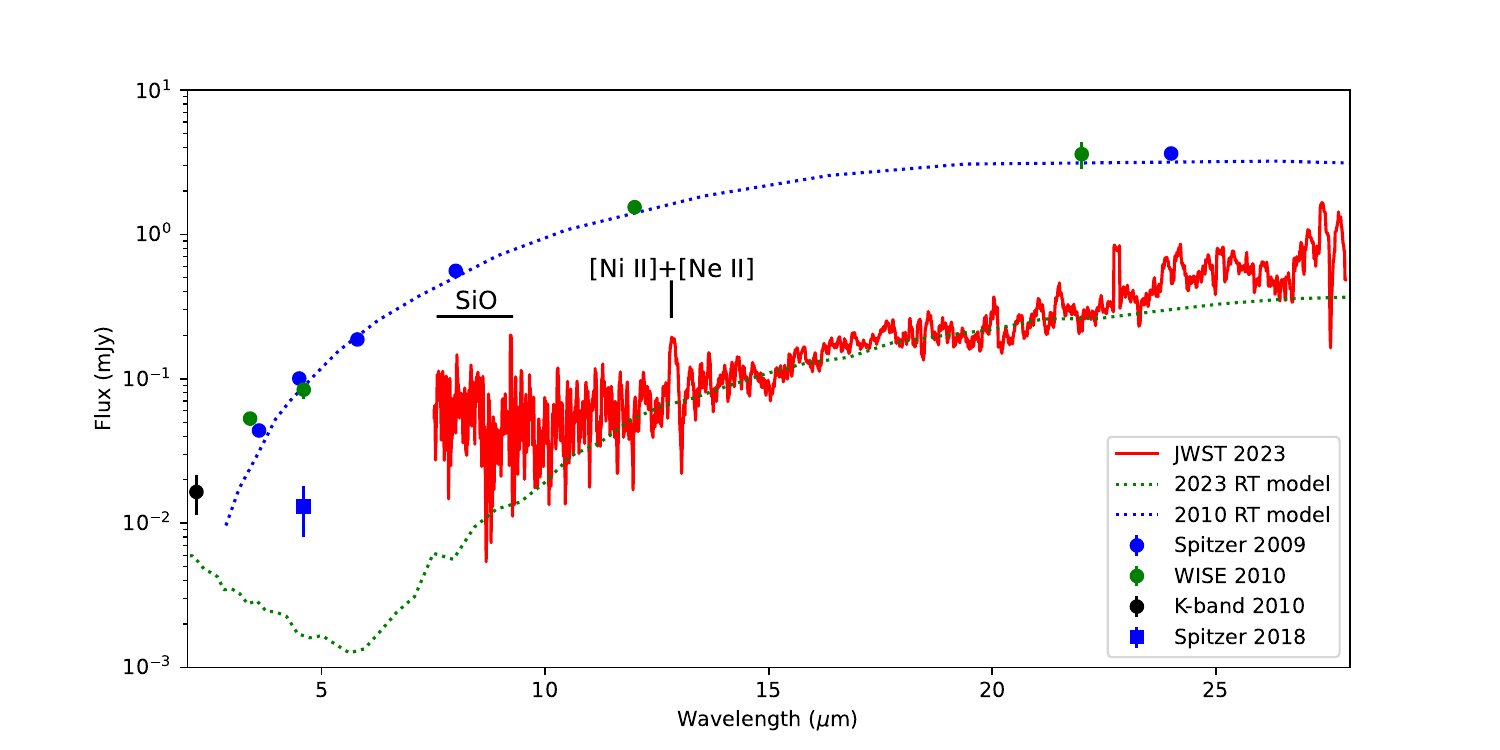}
\caption{The new {\it JWST}/MIRI spectrum is plotted along with the previous IR photometry. The best-fit radiative-transfer models to the 2009--2010 photometry and to the 2023 {\it JWST} spectrum are also plotted \citep{2023MNRAS.525.4928W}.}
\label{jwst-spec}
\end{figure*}

SN 1995N already showed a strong near-IR excess when first observed in 1996 July, about 1\,yr after the explosion, and observations at $t = 2$--7\,yr exhibit a large IR excess consistent with an IR echo \citep{2002ApJ...575.1007G}.
As shown in Figure~\ref{fig:lightcurve},
the SN was very bright at early times in the $K$ band, but mid-IR observations were not obtained of SN 1995N until 2009/2010 serendipitously
with the {\it Spitzer Space Telescope} and the {\it Wide-field Infrared Survey Explorer (WISE)}, $\sim$15\,yr after the explosion \citep{2013AJ....145..118V}. Figure~\ref{jwst-spec} shows the spectral energy distribution (SED) from the $K$ band through {\it Spitzer/MIPS} 24\,$\micron$ measured in 2009/2010.


SN 1995N was also detected by {\it Spitzer}/IRAC in the 4.5\,$\micron$ channel in 2018 \citep{2021ApJ...919...17S}; see Figure~\ref{jwst-spec}.
The authors note that at an age of $\sim 8600$\,d, it is the
latest observation of a SN IIn in this wavelength band.

The late-time near-IR luminosity can be explained by a simple model where a
small fraction of the optical and X-ray radiation is reprocessed into the IR by the interaction between preexisting
dusty CSM and the SN shock. As the shock moves away from the region
where the bulk of the CSM dust is located, the X-ray and IR dust emission fade away \citep{2005MNRAS.364.1419Z}.

The {\it NEOWISE} and Zwicky Transient Facility (ZTF) archives were searched but SN 1995N was not detected; however, it was 
visible in CTIO 4\,m/DECam images taken on 2014 May 26 ($g,r$ bands) and 2021 March 25 ($i$ band).
Differential photometry was done on these images, showing that SN 1995N was at $g \approx 23.2$ and $r \approx 22.1$\,mag in 2014, and $i\approx 23.1$\,mag in 2021. These data are plotted in Figure~\ref{fig:lightcurve}.

\subsection{Optical Spectra}
\begin{figure}
\centering
\includegraphics[width=3in]{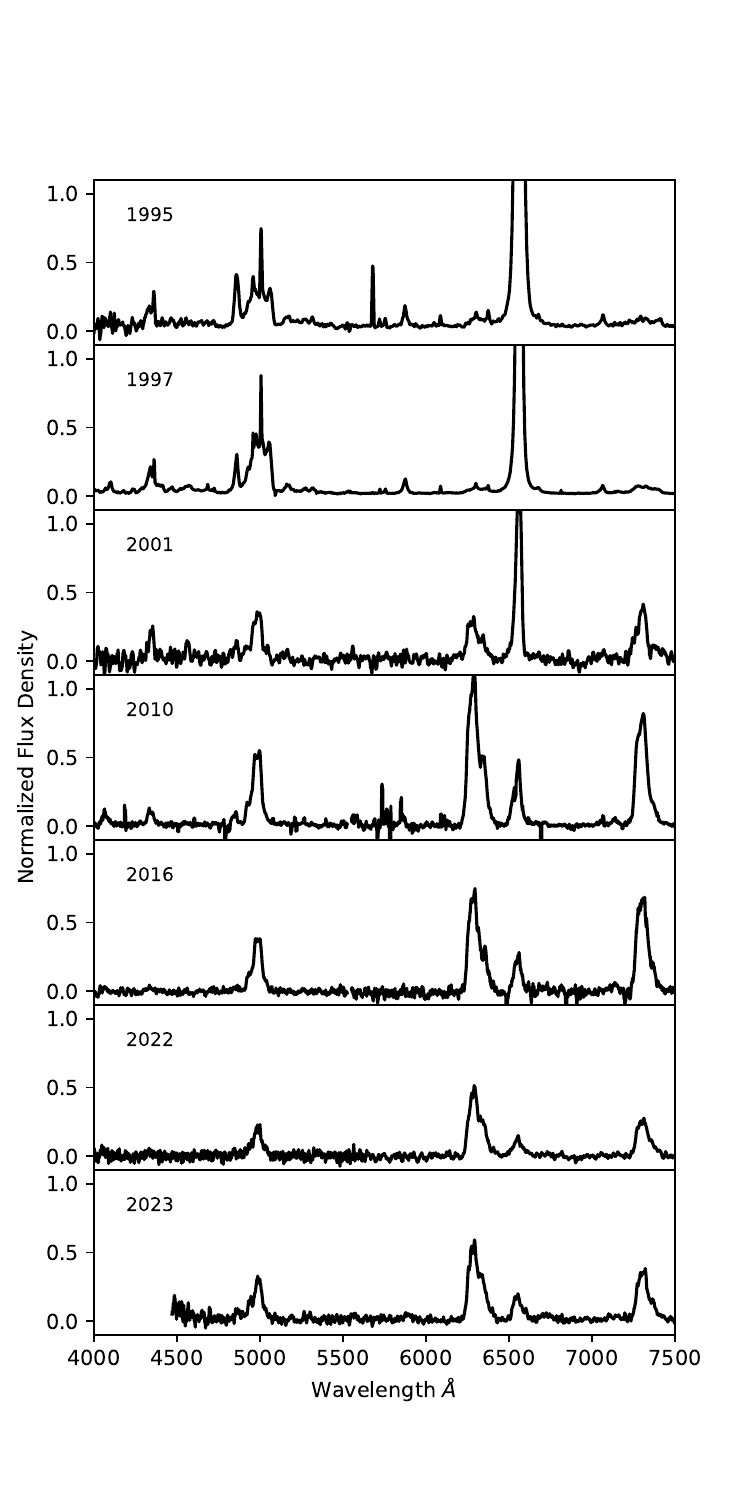}
\caption{Optical spectra of SN 1995N taken at various epochs from 1995 to 2023. The spectra have been smoothed using Astropy Box1DKernel to similar spectral resolutions. The flux densities ($f_\lambda$) have been normalized and shifted. See text for details. }
\label{fig:opticalspectra2}
\end{figure}

Two spectra of SN 1995N, obtained on 1995 July 28 and 2001 March 25 and downloaded from the Transient Name Server\footnote{https://www.wis-tns.org/object/1995n}, are plotted in Figure~\ref{fig:opticalspectra2}. Also plotted 
are two VLT/XShooter spectra taken in 2010 and 2016 \citep{2023MNRAS.525.4928W}. Not all of the spectra are flux calibrated, and they are all smoothed using Astropy Box1DKernel to similar spectral resolutions.

\begin{figure*}
\centering
\includegraphics[width=6in]{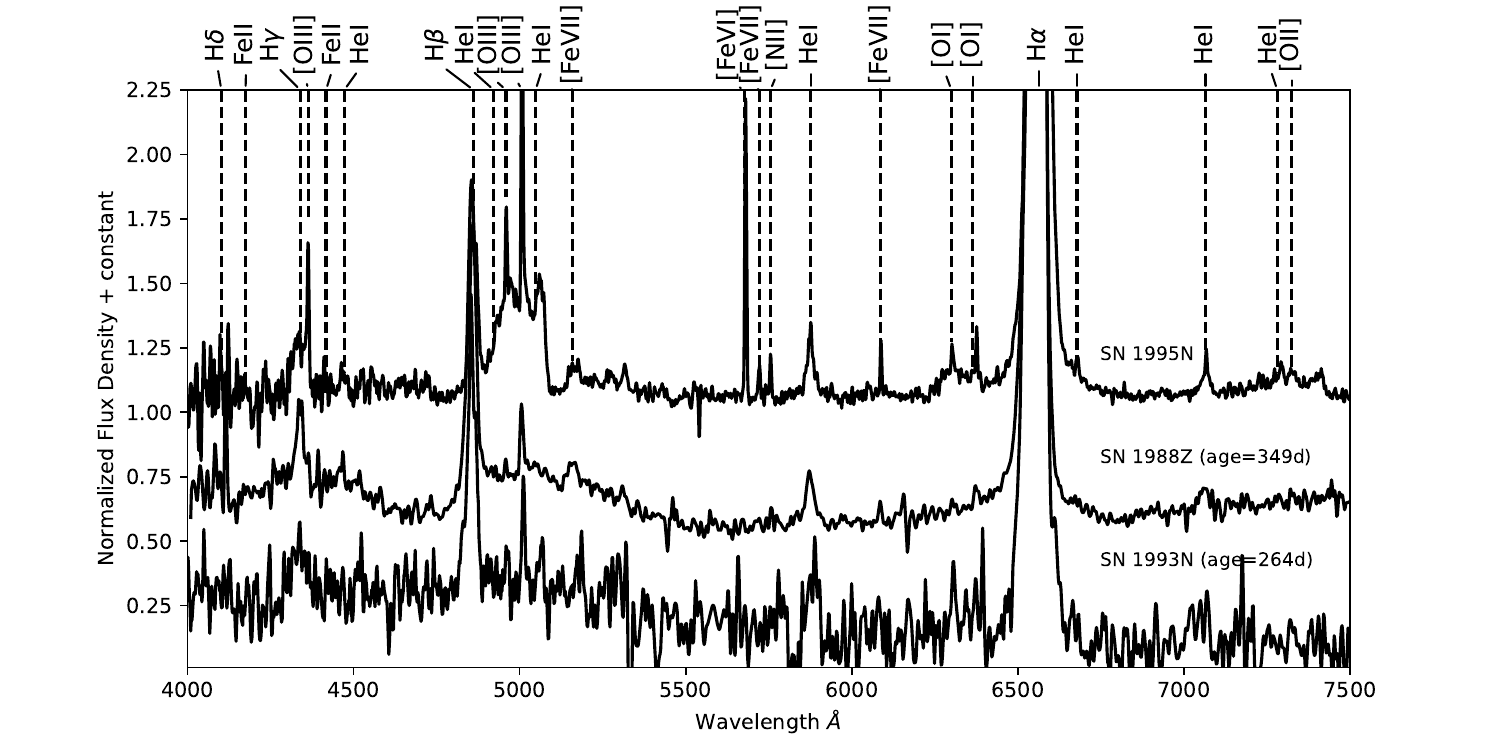}
\caption{An early-time optical spectrum of SN 1995N taken in 1995 July compared to early-time spectra of SN 1988Z and SN 1993N \citep{1991MNRAS.250..786S,1993MNRAS.262..128T}. Flux densities are $f_\lambda$.}
\label{fig:opticalspectra1}
\end{figure*}

Other early-time spectra including the discovery spectrum (1995 May 24) of SN 1995N are shown by \citet{1997ARA&A..35..309F} and  \citet{2002ApJ...572..350F}. 
The explosion date has been set to be 1994 July 4 (JD 2,449,537) assuming that the discovery spectrum, showing a single symmetric H$\alpha$  emission line, resembled that of SN 1993N about 10 months after the explosion \citep{1995IAUC.6170....1P, 2002ApJ...572..350F}. 
SN 1993N was discovered on 1993 April 15 \citep{1993IAUC.5784....1M}.
No spectra of SN 1993N have been published previously, but several early-time spectra were obtained with the Kast double spectrograph \citep{miller93} on the Lick Shane 3\,m telescope; one of them is shown in Figure~\ref{fig:opticalspectra1}.
The H$\alpha$ emission line has three components (narrow, intermediate, and broad), similar to SN 1988Z and SN 1995N, indicating CSM interaction \citep{1993IAUC.5788....1F,1993IAUC.5791....3M,1994IAUC.5924....1F}. Unlike these two SNe, SN 1993N was not detected at radio wavelengths \citep{2000MmSAI..71..331P}.

\citet{2023MNRAS.525.4928W} argue that the width of H$\alpha$ is not a good discriminator of age,  suggesting that the likely age at discovery is actually only $\sim$100 d.
However, \citet{1995IAUC.6170....1P} also note that coronal lines in the discovery spectrum of SN 1995N resemble those seen in SN 1988Z \citep{1993MNRAS.262..128T}. A SN 1995N spectrum (1995 July 7) taken two months after the discovery spectrum is plotted in Figure~\ref{fig:opticalspectra1} along with a spectrum of SN 1988Z taken at an age of $\sim 1$\,yr and a spectrum of SN 1993N taken at an age of $\sim 9$ months \citep{1991MNRAS.250..786S,1993MNRAS.262..128T}. The GELATO\footnote{https://gelato.tng.iac.es/} SN classification best fit actually picks an even later spectrum of SN 1988Z at day 1149 \citep{2008A&A...488..383H}. The SN 1995N discovery spectrum resembles the 1\,yr spectra of SN 1988Z, so the assumed age of 10 months at discovery is probably a lower limit \citep{1991MNRAS.250..786S,1993MNRAS.262..128T}.
No pre-discovery observations of SN 1995N have been found.

\begin{figure*}
\centering
\includegraphics[width=6in]{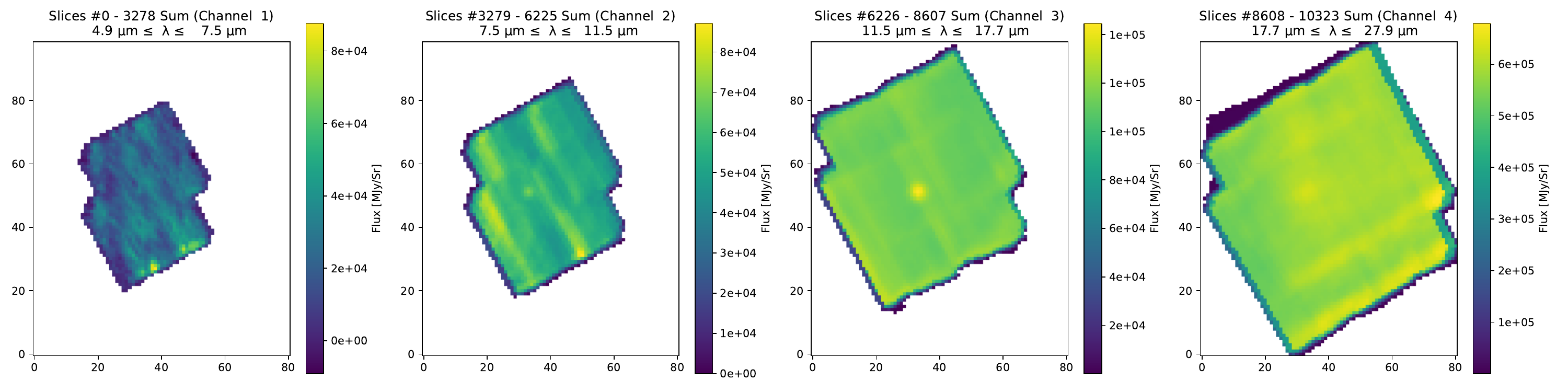}
\includegraphics[width=6in]{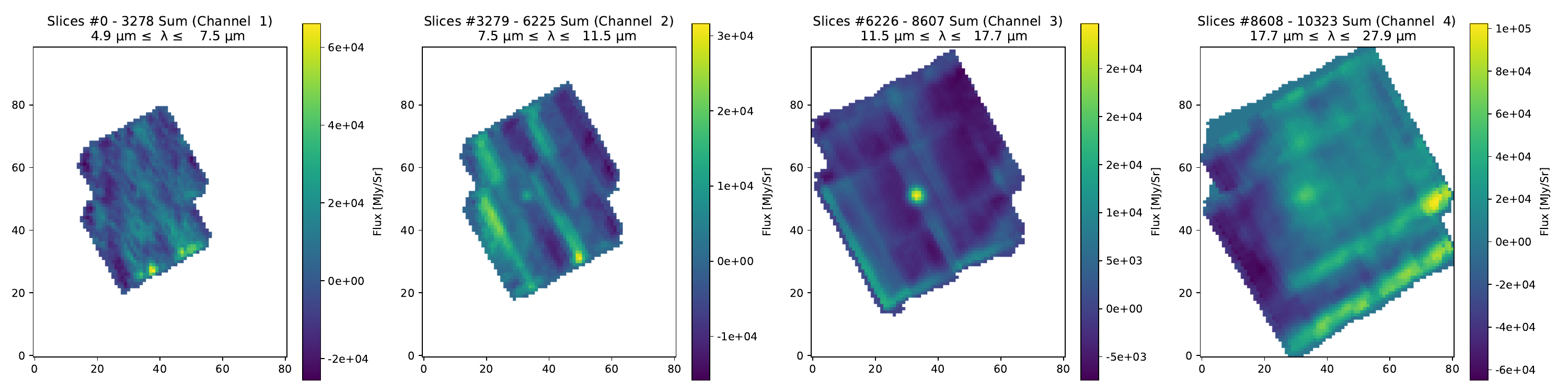}
\caption{Before (upper) and after (lower) background subtraction stacked images of the data cube in the 4 channels of MRS for SN 1995N. The emission from SN 1995N can be seen near the centers of all the cubes except in Channel 1.}
\label{fig:cube}
\end{figure*}

\begin{figure*}
\centering
\includegraphics[width=2in]{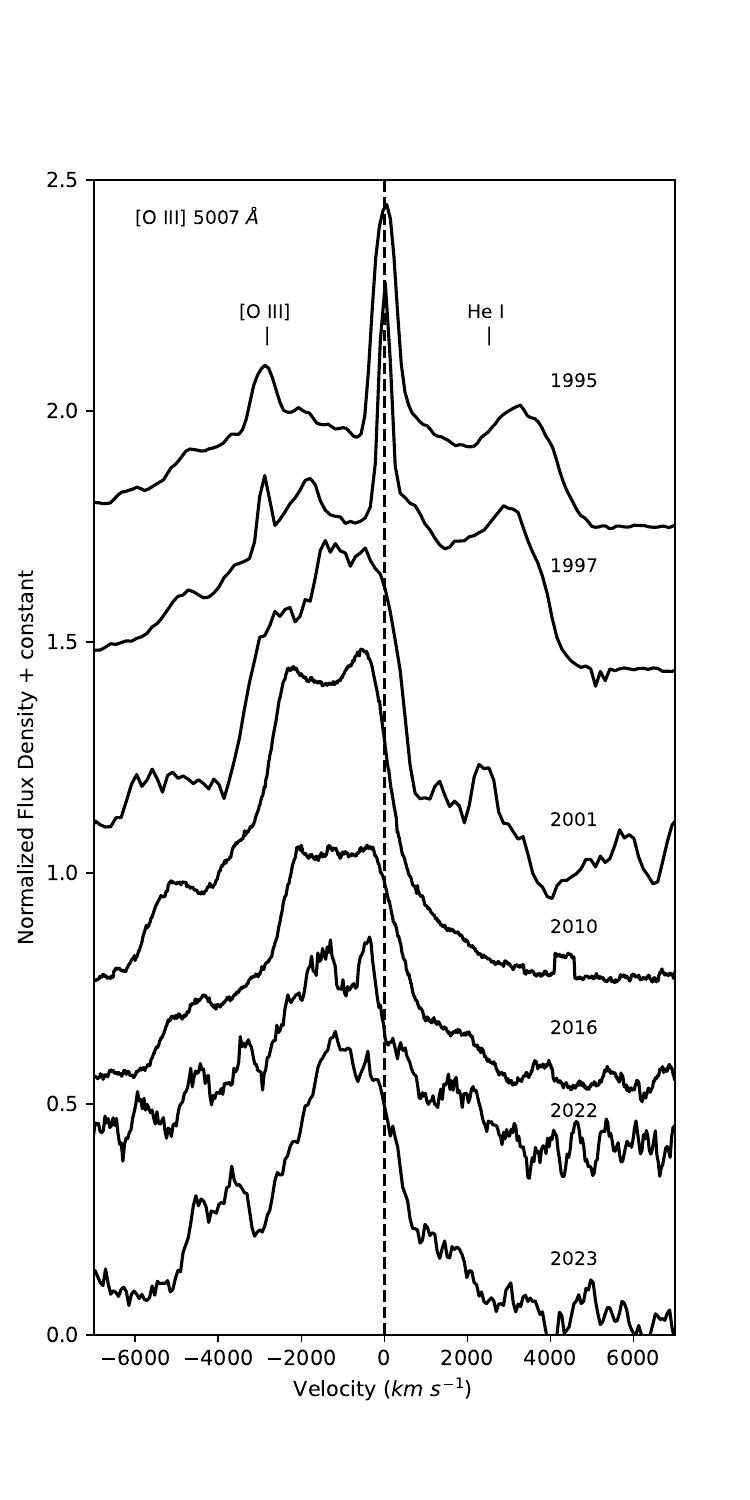}
\includegraphics[width=2in]{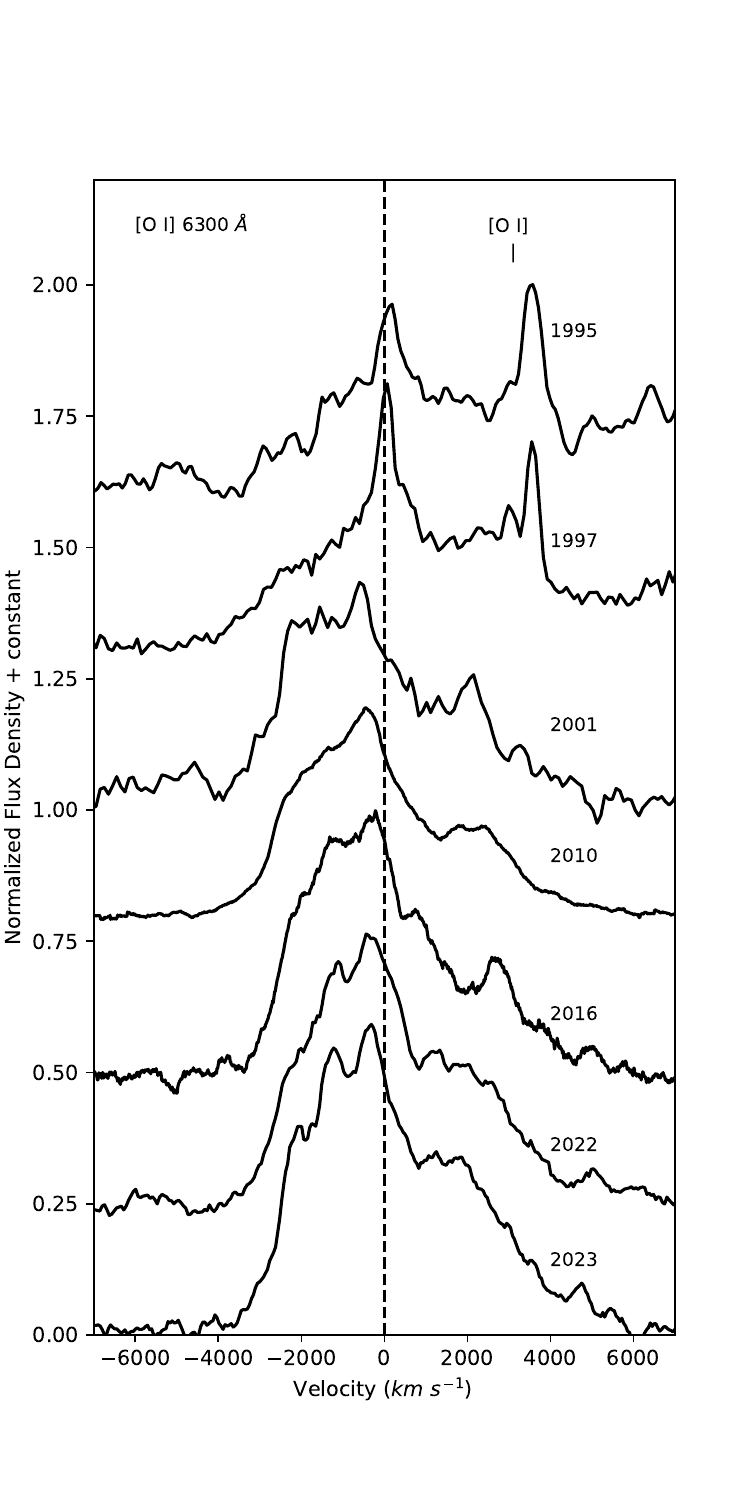}
\includegraphics[width=2in]{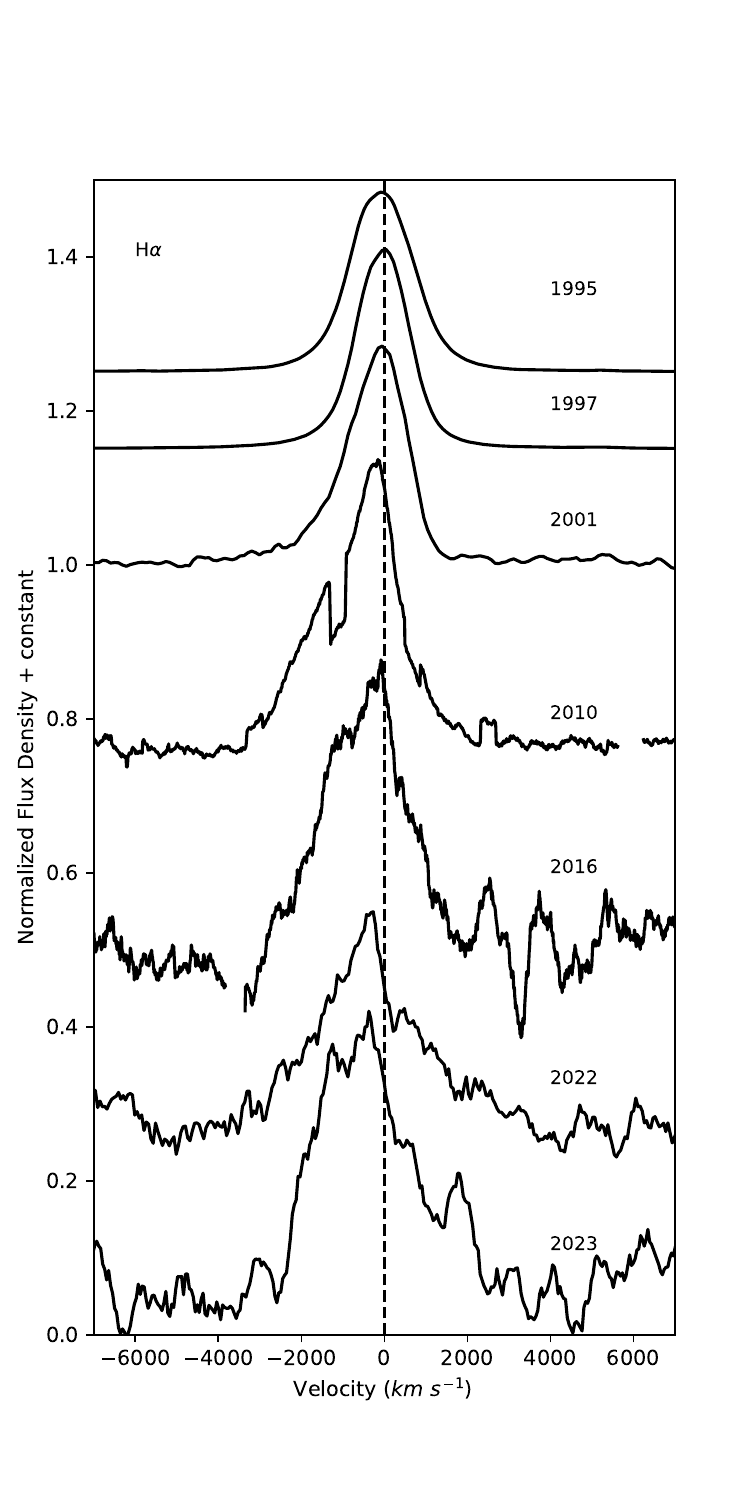}
\caption{Optical spectra of SN 1995N taken at various epochs from 1995 to 2023 in velocity space for [O III] 5007\,\AA, [O I] 6300\, \AA, and H$\alpha$ 6563\,\AA. The spectra were smoothed using Astropy Box1DKernel to similar spectral resolutions. The flux densities ($f_\lambda$)  have been normalized and shifted. See text for details. The rest wavelengths of [O III] 4959\,\AA, He~I 5048\,\AA, and [O~I] 6364\, \AA\ are marked.  }
\label{fig:opticalspectra3}
\end{figure*}


\section{New Observations}

New {\it JWST}/MRS observations of SN 1995N were obtained on 2023 July 2 (JD 2,460,129) with the {\it JWST}/MIRI MRS, 28\,yr (10,286\,d) since discovery.
The observations include all 4 channels
and all 3 wavelength ranges of the MRS. 

The data consist of $R = \lambda/\Delta\lambda = 1500$--3500 spectra spanning 4.9 to 27.9\,\micron, including all sub-bands. Dedicated offset background observations were also obtained to provide a more stringent estimate of the local thermal background.  
These observations are part of GO-1860 (PI O. Fox).
Figure~\ref{jwst-spec} shows the MIRI spectrum of the IR source at the position of SN~1995N as determined by 2MASS.   

Level 1 data were initially downloaded from MAST. The data were processed to Level 3 with version 1.11.1 of the {\it JWST} calibration pipeline and context jwst-1183 of the Calibration Reference Data System (CRDS). The standard MRS pipeline procedure was followed \citep{bushouse24}. This step automatically includes the dedicated background observations for subtraction.  
Figure \ref{fig:cube} shows the collapsed cube in each channel both before and after the dedicated background subtraction.

We note that the global thermal background is roughly 5 orders of magnitude larger than the SN 1995N flux; thus, estimating the local background is challenging. Even small variability in the background across the field of view can have a potentially large impact on the source spectrum, especially at the longest wavelengths where the background tends to be the largest and the signal-to-noise ratio (S/N) of the source tends to be the smallest.\footnote{https://www.stsci.edu/contents/news/jwst/2023/miri-mrs-reduced-count-rate-update} 

Furthermore, while the dedicated background removes the majority of the global thermal background, it does not take into account the local background of SN 1995N (see Figure \ref{fig:cube}). The background-subtraction method is described in detail by \citet{2024arXiv241009142S}.

Spectra of the SN were extracted using the {\it JWST} pipeline \texttt{extract\_1d} step.
For point sources, the pipeline uses a circular extraction aperture, which varies in radius with wavelength. The extraction-related vectors are found in the Advanced Scientific Data Format (ASDF) \texttt{extract1d} reference file. We use the default reference file and aperture correction.  Figure~\ref{jwst-spec} shows the final MRS extraction of SN 1995N.
Analysis of the data cubes shows that flux from the position of SN 1995N is visible starting around 7.6\,\micron, so the spectrum has been truncated below that wavelength.

Optical spectra were obtained with the Keck 10\,m telescopes on Maunakea using LRIS \citep{1995PASP..107..375O} on 2022 March 04 and DEIMOS \citep{2003SPIE.4841.1657F} on 2023 April 23.  
The LRIS observation used the 1\arcsec\ slit,
600/4000 grism, and 400/8500 grating. Data reduction followed standard techniques 
using the LPipe data-reduction pipeline \citep{2012MNRAS.425.1789S,2019PASP..131h4503P}. 
These spectra are plotted in Figure~\ref{fig:opticalspectra2}.

\section{Spectral Evolution}



Figures~\ref{fig:opticalspectra2} and \ref{fig:opticalspectra3} show the evolution of the optical spectra of SN 1995N over 28\,yr. The most obvious change with time is seen in H$\alpha$, whose line profile was quite symmetric at early times but has evolved with the red wing fading faster than the blue wing, an indication of dust formation in the ejecta.
The asymmetry in H$\alpha$ first becomes visible in a spectrum taken in 1999, 1799\,d past explosion \citep{2002ApJ...572..350F,2023MNRAS.525.4928W}.
The asymmetry in the hydrogen emission lines is also noted in a VLT spectrum taken on 2003 July 30 \citep{2005MNRAS.364.1419Z}.



In the VLT X-shooter spectra from 2010 and 2016, as well as the new Keck spectra from 2022/2023, the continued evolution can be seen in  Figures~\ref{fig:opticalspectra2} and \ref{fig:opticalspectra3}. 
When these spectra were obtained, SN 1995N was faint, $V \approx 23$\,mag. The optical brightness did not change much from 2010 to 2021, as shown in Figure~\ref{fig:lightcurve}.
Strong, broad emission from [O~I] $\lambda\lambda$6300, 6364, [O~II]
$\lambda\lambda$7319, 7331, and [O~III] $\lambda\lambda$4959, 5007 are seen in the spectra taken from 2010 to 2023. H$\alpha$ is weaker but also clearly detected. 
Between 2001 and 2010, the [O~I], [O~II], and [O~III] lines became much stronger relative to H$\alpha$. 
By 2010, the oxygen lines are far stronger than H$\alpha$, and this great strength has remained through 2023. In Figure 1 of \citet{2011AN....332..266P}, which shows the evolution of the optical spectrum of SN 1995N between 1995 and 2010, the strengthening of the [O~I] and [O~II] emission begins as early as 1998, and by 2004, these lines are nearly as strong as H$\alpha$.

\section{The JWST Spectrum}
The {\it JWST}/MRS spectrum is plotted in Figure~\ref{jwst-spec} along with {\it Spitzer} and {\it WISE} photometry to show how the IR emission has evolved over time. The emission feature seen at $\sim$13\,$\micron$ is probably a blend of [Ni~II]~12.729\,$\micron$ and  [Ne~II]~12.813\,$\micron$, seen strongly in the IR spectrum of SN 2004dj on day 868 \citep{2006ApJ...651L.117K,2011A&A...527A..61S}.
The [Ne~II] emission line is one of the primary coolants for the ejecta \citep{1992ApJ...395..540C}. The S/N is low but the plateau-like feature between 7.6 and 9\,\micron\ may be molecular emission due to the
SiO fundamental band as previously seen in SNe 1987a, 2003gd, 2004et, and 2005af \citep{1993ApJS...88..477W,2006ApJ...651L.117K,2007ApJ...665..608M,2012A&A...546A..28J}. \citet{1993ApJS...88..477W} found that the SiO feature disappeared in SN 1987A when dust formation occurred after day 400. This detection in SN 1995N, if real, would be at a much later time than previous detections in other SNe.




\section{Radiative-Transfer Modeling}
The dust associated with SN 1995N consists of  dust in the circumstellar shell that may have existed before the explosion and newly formed dust which may lie in the unshocked
ejecta not yet hit by the reverse shock, or in the
cool dense shell between the forward and reverse shocks \citep{2023MNRAS.525.4928W}.

Before 2009, there were no IR observations of SN 1995N to the red of the $K$ band. The large excess in the $K$-band photometry undoubtedly results from emission by warm dust but cannot be used to estimate dust masses \citep{Gerardy_2002}.

\citet{2013AJ....145..118V} modeled the 2009/2010 dust emission detected using {\it Spitzer}/IRAC at 3.6, 4.5,
5.8, and 8.0\,$\micron$ and {\it Spitzer}/MIPS at 24\,$\micron$, as well as with {\it WISE} at 3.4, 4.6, 12, and 22\,$\micron$.
They used an ``idealized dust cloud," which is just an optically thin point source, to model the dust emission from the {\it Spitzer/WISE} photometry. These models were calculated using astronomical silicate and graphite grains with a size of 0.1\,$\micron$.
The model results imply a dust mass of 0.05\,M$_{\sun}$ for silicate grains and 0.12\,M$_{\sun}$ for graphite grains.  \citet{2013AJ....145..118V} suggest that the dust is cool, $\sim 240$\,K, and likely is pre-existing dust from circumstellar mass loss before the explosion. 

\begin{figure*}
	\centering
    \includegraphics[width=3in]{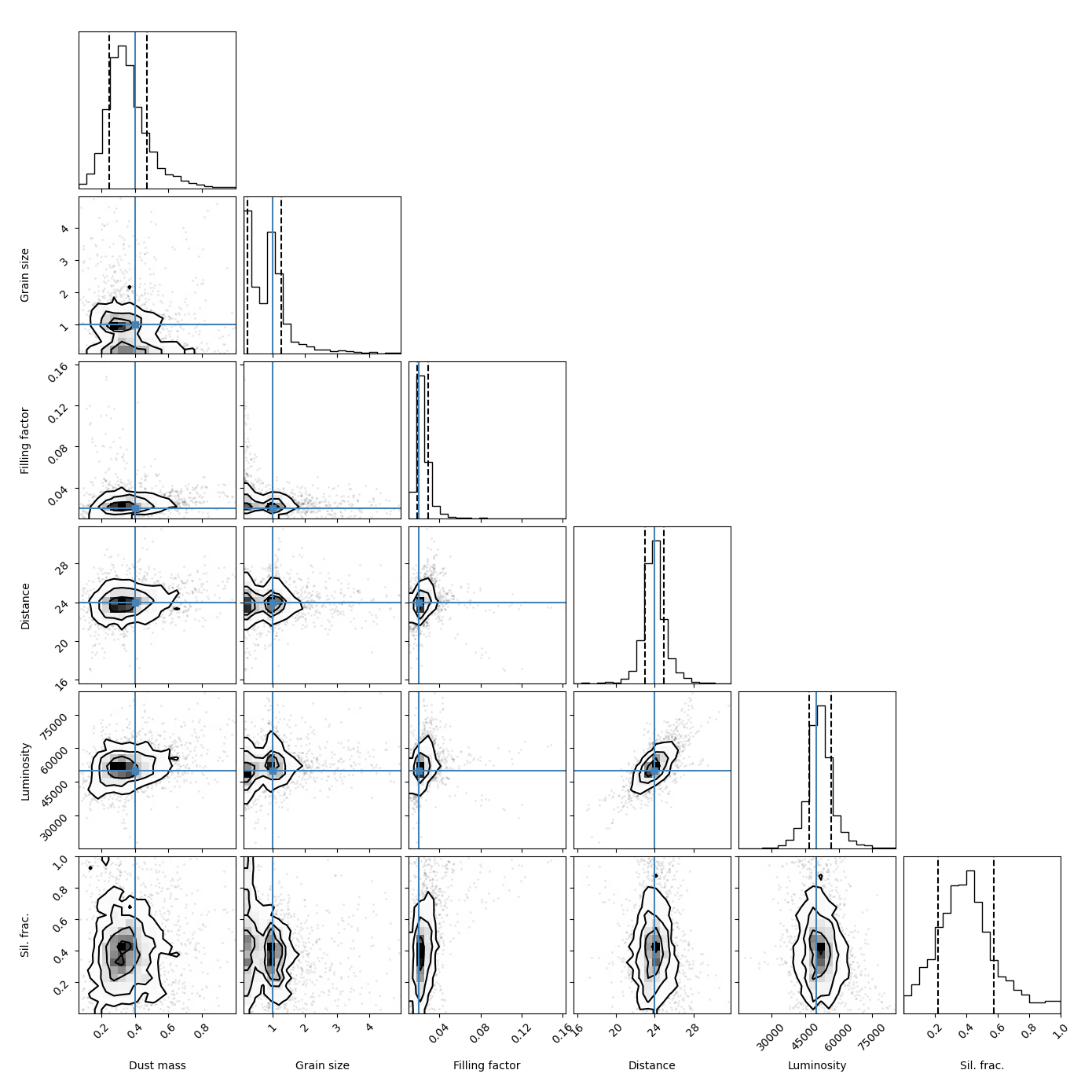}
	\includegraphics[width=3in]{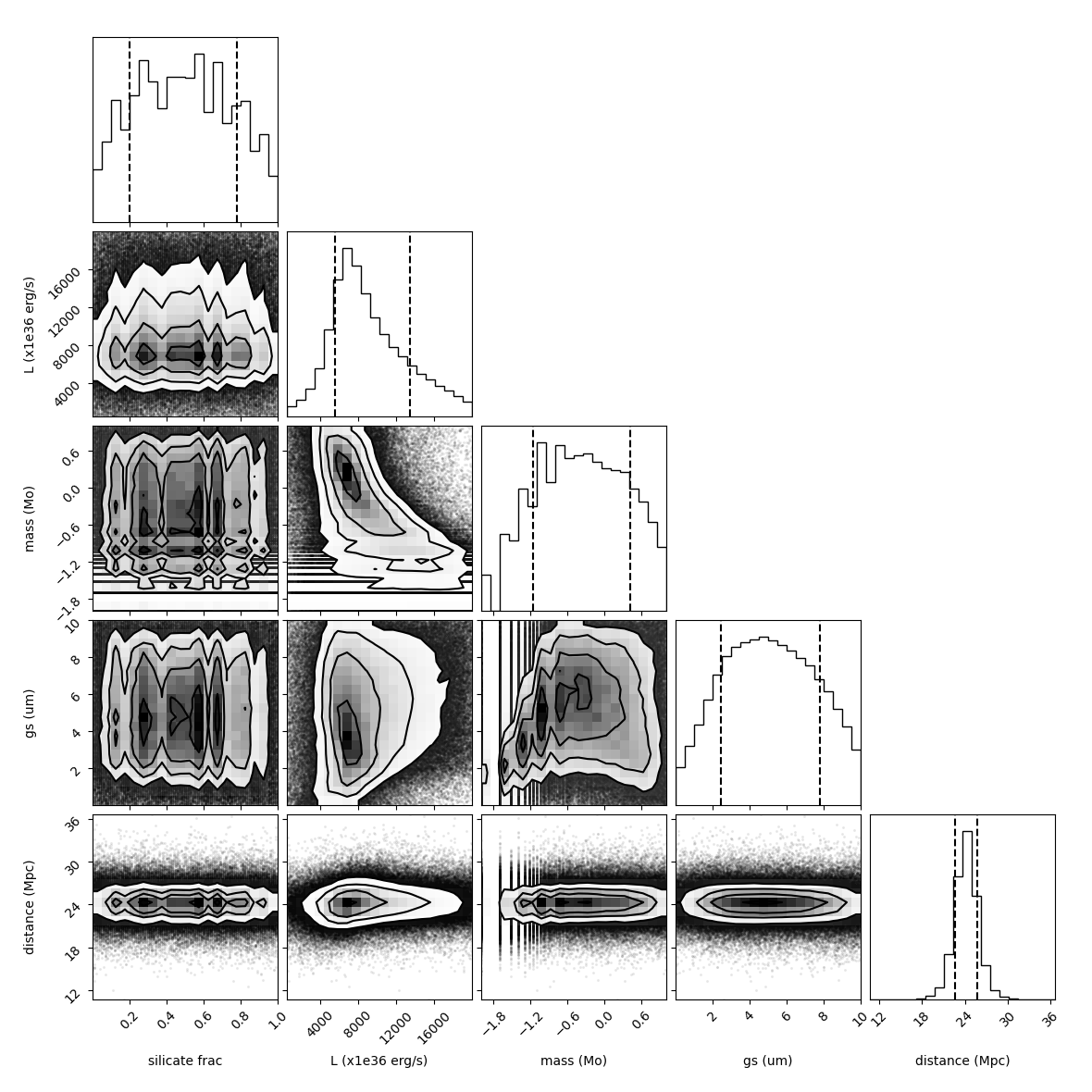} 
		\caption{The results of the MCMC analysis of the {\sc mocassin} models for the {\it Spitzer/WISE} 2009-10 epoch on the left and for the {\it JWST} 2023 epoch on the right. The parameters are silicate fraction, heating source luminosity, dust mass, grain size, and distance. Vertical dashed lines indicate 1$\sigma$ confidence intervals. Blue crosshairs in the left panel indicate the values found by \citet{2023MNRAS.525.4928W} from an analysis of a much smaller grid of models than used here.}
\label{fig:mcmc}
\end{figure*}

\citet{2023MNRAS.525.4928W} remodeled the {\it Spitzer} and {\it WISE} data used by \citet{2013AJ....145..118V}. 
They employed the three-dimensional (3D) radiative-transfer code {\sc mocassin} \citep{2005MNRAS.362.1038E}  and found a best-fit model with 0.4\,M$_{\sun}$ of 1\,$\micron$ amorphous carbon dust
grains, in a clumpy shell with a volume filling factor of 0.02. Figure~\ref{fig:mcmc} shows the MCMC corner plot for these models. The parameters are listed in Table~\ref{tb_mocassin}, and
this fit is plotted in Figure~\ref{jwst-spec}. 

Optical spectra presented by \citet{2002ApJ...572..350F} and VLT Xshooter spectra obtained in 2010 and 2016 were also modeled using {\sc damocles} \citep{2016MNRAS.456.1269B}, a code which calculates the profiles of emission lines arising within the expanding ejecta including the effects of dust embedded in the ejecta \citep{2023MNRAS.525.4928W}. The spectra which were fit are plotted in Figure~\ref{fig:opticalspectra2}.
The \citet{2002ApJ...572..350F} spectrum from 1999 (JD 2,451,126) was modeled and a dust mass of $2 \times 10^{-4}$\,M$_{\sun}$ was found.
The best-fit dust masses for the 2010 and 2016 spectra estimated with this alternative emission-line method were consistent with the {\sc mocassin} results of 0.4\,M$_{\sun}$ for the 2009/2010 IR epoch but are also well fit with dust masses greater than 0.1\,M$_{\sun}$. 

New {\sc mocassin} models were constructed to fit the {\it JWST} data presented here. As a starting point, we took the best-fit model grid and clump parameters from \citet{2023MNRAS.525.4928W}, and then expanded the grid to the epoch of the {\it JWST} spectra to take into account the expansion of the ejecta between 2010 and 2023. That best-fit model has 0.4\,M$_{\sun}$ of amorphous carbon dust in the form of 1\,\micron\ grains, distributed in a clumpy shell with a volume filling factor of 0.02. The grid which fit the observations in 2009--2010 was enlarged by assuming constant homologous expansion. Models with all parameters unchanged except the grid size significantly overpredict the {\it JWST} fluxes. The heating source has therefore significantly declined in luminosity since 2009--2010.

Therefore, models were constructed with an expanded grid and a reduced source luminosity. These can broadly fit the {\it JWST} observations. Thus, a scenario in which dust formation was complete by 2009--2010, and the dust existing at that epoch has since expanded and cooled, may be able to account for the observed SED in 2023.

To assess the range of parameters which can fit the {\it JWST} data, we explored a much larger range of parameters using the Markov Chain Monte Carlo (MCMC) ensemble sampler \texttt{emcee} \citep{2013PASP..125..306F}. Approximately 320,000 models were run to sample the parameter space which we describe briefly here.

\noindent $\bullet$ Distance: we take a Gaussian prior with a mean of 24.1\,Mpc and standard deviation of 1.6\,Mpc (taken from the NASA Extragalactic Database, and consistent with that adopted by previous studies; \citealt{2002ApJ...572..350F, 2013AJ....145..118V}).

\noindent $\bullet$ Heating source luminosity:  the dust in the models is heated by a diffuse source distributed uniformly in the interclump regions of the expanding dust shell. Physically, this is consistent with a scenario in which the energy source powering the dust emission is the interaction between the ejecta and CSM, but rather than directly heating the dust, this interaction heats gas within the ejecta, which in turn heats the dust. The luminosity of the heating source will depend on the extent of the ejecta-CSM interaction, and has clearly declined significantly since the 2009--2010 \textit{Spitzer}/{\it WISE} data. We adopt a uniform prior with value of (5--200) $\times 10^{36}$\,erg\,s$^{-1}$.

\noindent $\bullet$ Grain size: our models assume a single grain size. A distribution of grain sizes is more realistic, but would require at least two more free parameters (upper and lower limits and power-law exponent for an MRN-type grain-size distribution; \citealt{1977ApJ...217..425M}) on which few observational constraints are available. We therefore use only a single grain size, with an assumed uniform prior from 0.005--10\,$\micron$.

\noindent $\bullet$ Silicate fraction: \citet{2023MNRAS.525.4928W} argue that silicate dust was unlikely to be present within the ejecta in 2009--2010, owing to the absence of large red scattering wings in optical emission lines. Nevertheless, we consider the possibility that some silicate dust has subsequently formed, and assume a uniform prior from zero to unity of the silicate fraction.

\noindent $\bullet$ Dust mass: we adopt a broad and uniform prior in log space for the dust mass, extending from $10^{-2}$ to $10^1$\,M$_{\sun}$



The {\it JWST} spectrum was resampled to match the sampling of the {\sc mocassin} output SED, and
the likelihood function was taken as the sum of the $\chi^2$ values obtained by comparing model fluxes to the observations. 200 walkers were used for 2500 iterations to explore the parameter space and estimate the posterior probability distributions.


\begin{figure*}
	\centering
\includegraphics[width=\textwidth]{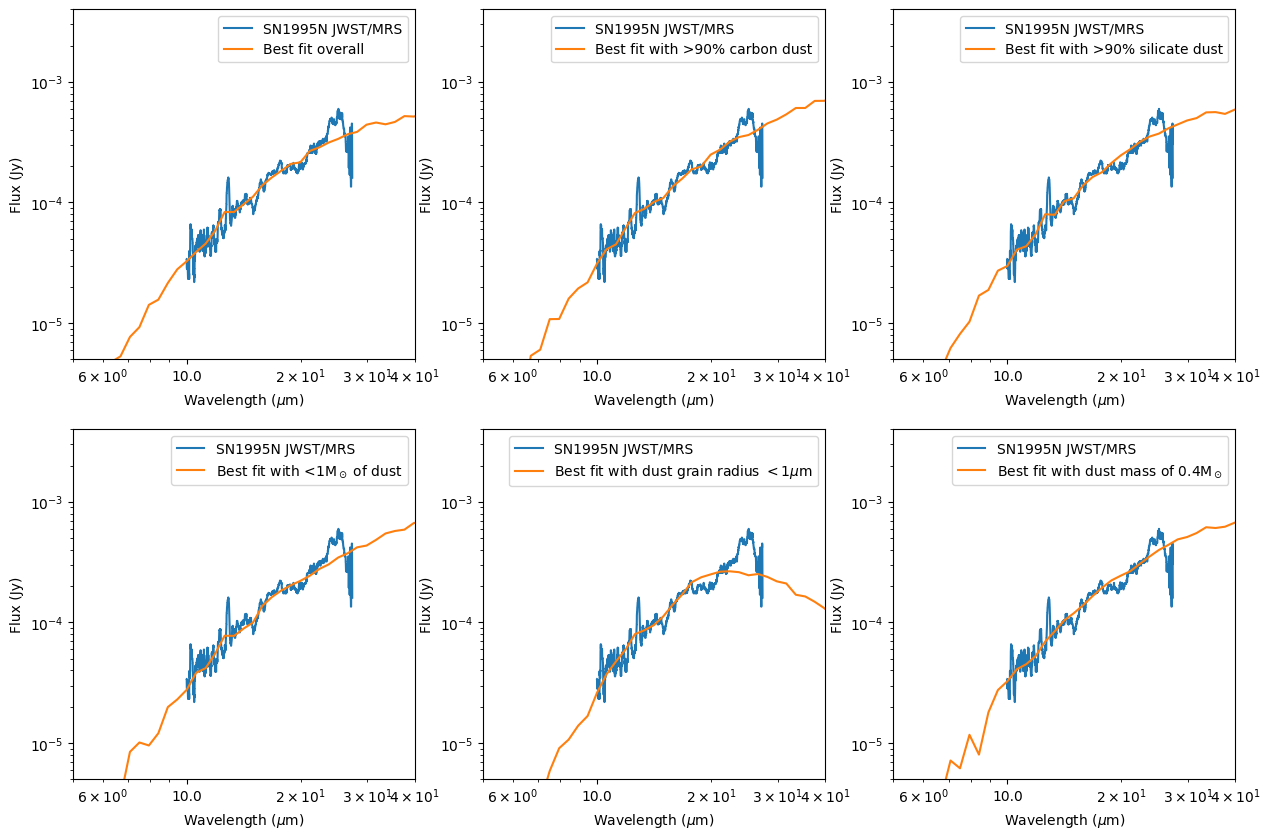}
		\caption{Best-fit {\sc mocassin} models for the new {\it JWST} spectrum; see Table~\ref{tb_mocassin}.}
\label{fig:mocassin}
        
\end{figure*}

The results of the MCMC analysis of the parameter space are shown in Figure~\ref{fig:mcmc}. 
The heating source luminosity is well constrained at $\sim 8 \times 10^{39}$\,erg\,s$^{-1} = 2 \times 10^6$\,L$_\odot$, given the assumptions for the distance. The silicate fraction is not well constrained by the data. At very late times, the dust in SN ejecta can be cold enough, or the grains large enough, that the SED of silicate dust is quite featureless \citep{2010ARA&A..48...21H}.
Dust masses below 0.1\,M$_{\sun}$ are strongly disfavored,
and dust grains several microns in radius are preferred. Dust masses of $\sim 1$\,M$_{\sun}$ give the best fits, but they are not well constrained either. Lower dust masses require higher luminosities and smaller grains. 

The best fits for various parameter values are shown in Figure~\ref{fig:mocassin}, and the parameters of these models are given in Table~\ref{tb_mocassin}. Figure~\ref{fig:mocassin} demonstrates how similar many of the fits are, and although the $\chi^2$ values listed in Table~\ref{tb_mocassin} differ, the fits appear very similar to the eye.

\begin{table*}
	\caption{Best {\sc mocassin} Models}
	\def\arraystretch{1.15}
	\begin{tabular}{lcccccc}
		\hline
Model&$\chi^2$& sil. frac.& $L$ ($10^{39}$\,erg\,s$^{-1}$)& $M_d$ (M$_{\sun}$)& grain size ($\micron$)& distance (Mpc)\\      
		\hline
{\it Spitzer/WISE } models&\nodata&0.405$\pm$0.192& 51.8$\pm$6.3&0.36$\pm$0.13& 0.9$\pm$0.7&24.0$\pm$1.3\\
\hline
JWST models:&&&&&&\\
$M_d$ = 0.4\,M$_{\sun}$ &9230.6&0.05&9.0&0.40&3.8&27.5 \\
Best of all fits& 4031.2&0.43&6.1&2.62&6.2&25.1\\
$>$90\% silicates&4776.2&0.94&6.0&3.64&8.1&24.3\\
$>$90\% carbon& 4896.2&0.01&6.7&0.82&4.4&24.0\\
$M_d < 1.0$\,M$_{\sun}$ &5371.5&0.11&7.3&0.87&4.1&26.6\\
grain size$< 1$\,\micron &9194.6  & 0.77  &  5.4    &    0.04  &  0.9  & 23.4\\
\hline
\hline
	\end{tabular}
 

	\label{tb_mocassin}
\end{table*}

As can be seen for the examples in Table~\ref{tb_mocassin}, many of 
the best fits have dust masses significantly greater than the 0.4\,M$_{\sun}$ inferred from the modeling of the 2009/2010 photometry.
This suggests that dust formation in SN 1995N could be continuing between 2009--2010 and 2023, although some of the inferred dust masses are unprecedented. \citet{2023MNRAS.525.4928W} found that a grain size of 1\,$\mu$m gave the best fit to the 2009--2010 SED, while the current analysis suggests larger grains still, pointing  to possible accretion onto previously formed grains as a likely mechanism for the continuing dust-mass growth.



\section{The evolution of SN 1995N}

\citet{2002ApJ...572..350F} found that the velocities and
densities measured from the narrow lines in the early-time
spectra of SN 1995N were typical for the CSM of red supergiants.  
They also suggested that the SN 1995N progenitor
was similar to the highly luminous red supergiant VY~CMa and the post-red-supergiant IRC~+10420, which have superwinds and initial masses
of at least 30\,M$_{\sun}$. This kind of extreme mass-loss rate is required to account for the dense CSM and strong SN-CSM interaction seen in SN 1995N. 
The dust-emitting volume of SN 1995N, $r\approx10^4$\,AU, is compatible with the size of the CSM around VY~CMa and IRC~+10420, and the estimated mass of the SN 1995N progenitor nebula agrees with those found for the two Galactic stars \citep{2013AJ....145..118V}.

Optical and ultraviolet spectra of SN 1995N, obtained 1--5\,yr after the explosion, indicate the presence of three velocity components with narrow (500\,km\,s$^{-1}$), intermediate (5000\,km\,s$^{-1}$), and broad (10,000\,km\,s$^{-1}$) line widths \citep{2002ApJ...572..350F}. 
Most of the gas is in the intermediate component but weak wings extend out to 10,000\,km\,s$^{-1}$. The velocities of the lines for SN~1995N are not as large as in SN 1979C, for example, suggesting a weaker explosion.

The emission from SN 1995N comes from clumps in the X-ray-photoionized, preshock, circumstellar gas (narrow lines), shocked gas (intermediate lines), and SN ejecta freely expanding  or possibly interacting with clumpy or asymmetric CSM (broad lines) \citep{2002ApJ...572..350F}.  It is likely that all the emission
components are heated and ionized by high-energy radiation from the interaction region.
These mechanisms were also suggested to explain the observations of SN 1988Z \citep{1994MNRAS.268..173C}.

The later-time evolution of the SN 1995N spectrum can be seen in 
Figures~\ref{fig:opticalspectra2} and \ref{fig:opticalspectra3}. The most obvious change with time is in H$\alpha$, whose line profile was quite symmetric at early times but starting in 1999, about 5\,yr after the explosion, evolved with the red wing fading faster than the blue wing, an indication of dust formation in the ejecta \citep{2002ApJ...572..350F,2023MNRAS.525.4928W}. 

To get to the observer, redshifted photons from the receding far edge of the ejecta must pass through the dust mixed with the expanding ejecta while the blueshifted photons from the approaching near edge do not, resulting
in more attenuation of the red wing of intrinsically symmetric emission lines.
There is also dust in the preexisting CSM,  but the high velocities shown here indicate dust has formed in the ejecta. 
This is clearly visible in Figure~\ref{fig:opticalspectra3} which shows how the [O III], [O I], and H$\alpha$ emission lines evolve in velocity space.
Also, a blue/red asymmetry clearly developed after 1--2\,yr in the intermediate-width H$\alpha$ component, 
suggesting that some of the new dust is in the cool dense shell behind the shock.

The SN-CSM interaction began sometime before the discovery spectrum was obtained $\geq$10 months after the explosion. SN 1995N was very bright at X-ray and radio wavelengths, indicating a strong ejecta-CSM interaction. The X-rays produced by the interaction photoionized hydrogen in the CSM shell lead to H$\alpha$ emission.
The SN–CSM interaction also caused
a reverse shock to form that moved back into the expanding ejecta \citep{2017hsn..book.2211M}. Eventually, this reverse shock reached the inner ejecta region, ionizing the metal-rich gas from which the optical
lines are produced. This caused the [O I], [O II], and [O III] emission lines to strengthen until they became stronger than H$\alpha$.
This is part of the transition from SN to SNR, when the radioactive $^{56}$Co heating of the O-rich inner ejecta gives way  to the reverse-shock excitation \citep{2017hsn..book.2211M}.
Such an abrupt change in emission is rarely observed in SNe,
but it is an important step in the transition from SN to SNR. It also implies that the SN 1995N
progenitor was fully or partially  stripped of its
H-rich envelope before the explosion.

A similar metamorphosis where the O emission becomes stronger than the H emission is seen in late-time spectra of SN 1996cr \citep{2024arXiv241213024P}. Two other SNe, SN 2001em
and SN 2014C, exploded as stripped-envelope SNe but later interacted with H-rich
circumstellar shells, changing them from Type Ib/c to IIn \citep{2006ApJ...641.1051C,2015ApJ...815..120M}.
Neither SN 1995N nor SN 1996cr have  spectra taken near the time of the explosion, so the early Type Ib/c phase may have been missed.
Both SN 1996cr and SN 1995N underwent a major episode of mass loss before their explosions.


The dust associated with SN 1995N includes pre-existing dust in the CSM which is heated by the initial flash and then by the ejecta-CSM interaction. The dust emission is seen in $K$-band emission early, but no estimate of the dust mass can be made because {\it Spitzer/WISE} photometry is not available until 2009/2010, 14\,yr after the explosion. The dust formation in the ejecta began in the late 1990s. Dust can also form in the cool dense shell behind the forward shock. The dust measured in the 2009/2010, 2018, and 2023 IR observations 
could be in the CSM, ejecta, or cool dense shell; these alternatives cannot be distinguished. 

The {\sc damocles} code \citep{2016MNRAS.456.1269B}, which fits [O I], [O III], and H$\alpha$ emission-line profiles in the optical, estimates only the mass of dust in the ejecta \citep{2023MNRAS.525.4928W}. {\sc damocles} was used to fit optical spectra of SN 1995N obtained between 1996 and 2016 \citep{2002ApJ...572..350F,2023MNRAS.525.4928W}. Assuming amorphous carbon dust, the dust mass in the ejecta from 1996 to 1999 is $\lesssim$10$^{-4}$ M$_{\sun}$. For spectra obtained in 2010 and 2016, the dust mass is $\gtrsim$0.1 M$_{\sun}$.
In particular, the 2010 {\sc damocles} estimate and the 2009/2010 {\sc mocassin} estimate are both compatible with a dust mass of $\sim$0.4 M$_{\sun}$. So, most or all of the dust mass in that epoch could be embedded in the ejecta and formed sometime after 1999, about 4\,yr after the explosion. There is no IR photometry until $\sim15$\,yr after the explosion, so a comparison with the emission-line dust-mass estimates is not possible at early times. 

The dust mass in 2009/2010 estimated through {\sc mocassin }radiative-transfer modeling is 0.36$\pm$0.13 M$_{\sun}$ \citep{2023MNRAS.525.4928W}. The estimated dust mass in 2023 using the {\it JWST} spectrum is 0.42$^{+2.16}_{-0.35}$  M$_{\sun}$.

\section{Conclusions}

The early-time spectra of SN 1995N exhibit narrow, intermediate, and broad emission lines, arising respectively from the photoionized CSM shell, the interaction of the CSM lost by the star before the explosion with the ejecta, and the freely expanding ejecta. The narrow lines imply a dense, massive CSM shell resulting from a large red-supergiant mass-loss episode shortly before the SN explosion. This mass loss partially or fully stripped the envelope, so SN 1995N could have been intrinsically a Type Ib/c SN shortly after the explosion, but the CSM interaction began before the discovery spectrum was obtained nearly a year later. The result is a very unusual late-time spectrum where the oxygen forbidden emission lines are much stronger than H$\alpha$. This occurs when the reverse shock begins to interact with and excite the O-rich, H-poor inner ejecta.
The spectral evolution indicates that the transition from SN to SNR has begun \citep[e.g.,][]{2008ApJ...677..306M}.

The dust associated with SN 1995N may be in the preexisting CSM, the expanding ejecta, or the cool dense shell behind the forward shock. The IR observations measure the total dust mass in these three locations while the emission-line fitting measures only the dust forming in the ejecta. 

As shown in Figure~\ref{fig:mcmc}, the dust shell inferred by \citet{2023MNRAS.525.4928W} can account for the {\it JWST} spectrum without difficulty if it has uniformly expanded in the meantime and the heating source has faded. While a fit is possible with many different combinations of parameters, the whole ensemble suggests that both the dust mass and grain size have increased between 2010 and 2023.

However, these are the best models that can be produced considering that the data only cover $\sim$7 to 27\,\micron. The estimate of the dust mass is not well constrained in the absence of data longward of 27\,\micron. When one looks at Figure~\ref{fig:mocassin}, the differences among the fits are quite small. The model with the same dust mass as in 2010, 0.4\,M$_{\sun}$, along with the expansion of the ejecta and fading of the heating source over the last decade, is perhaps the most likely using Occam's razor. A dust mass of 0.4\,M$_{\sun}$ is also more consistent with the masses observed in other older CCSNe. A large mass of cold dust such as that seen in SN 1987A would not be detected in these observations \citep{2011Sci...333.1258M,Matsuura_15}.

As shown in Figure 16 of \citet{2023MNRAS.525.4928W}, SN 1995N is one of the oldest Type II SNe where IR dust emission and optical emission lines can still be detected, over 10,000 days 
post-discovery.
More observations of CCSN dust emission at extremely late times are needed that extend further into the IR. Until then, it will be difficult to choose between the two dust-formation scenarios --- continuous dust formation or most (or all) of the dust forms at early times following the SN explosion \citep{2015ApJ...810...75D,2016MNRAS.456.1269B,2019ApJ...871L..33D,2023MNRAS.525.4928W}.

Similarly, the very late-time optical spectra allow a rare view into the inner ejecta excited by the reverse shock, showing the transition from SN to SNR.

\begin{acknowledgments}
This work is based (in part) on observations made with the NASA/ESA/CSA {\it James Webb Space Telescope}. The data were obtained from the Mikulski Archive for Space Telescopes at the Space Telescope Science Institute, which is operated by the Association of Universities for Research in Astronomy, Inc., under NASA contract NAS 5-03127 for {\it JWST}. The observations are associated with program GO-1860.

Partial support for this work was provided by NASA grant JWST-GO-02666.002-A.
S.Z. received support from the NKFIH OTKA K142534 grant.
A.V.F.’s research group at UC Berkeley acknowledges financial assistance from the Christopher R. Redlich Fund, as well as donations from Gary and Cynthia Bengier, Clark and Sharon Winslow, Alan Eustace and Kathy Kwan, William Draper, Timothy and Melissa Draper, Briggs and Kathleen Wood, and Sanford Robertson (W.Z. is a Bengier-Winslow-Eustace Specialist in Astronomy, T.G.B. is a Draper-Wood-Robertson Specialist in Astronomy, Y.Y. was a Bengier-Winslow-Robertson Fellow in Astronomy), and many other donors.

The W. M. Keck Observatory is operated as a scientific partnership among the California Institute of Technology, the University of California, and NASA; the observatory was made possible by the generous financial support of the W. M. Keck Foundation. The Kast spectrograph on the Shane 3\,m telescope at Lick Observatory was made possible through a generous gift from   William and Marina Kast. We acknowledge the excellent assistance of the staff at each of Keck and Lick Observatory. 

\end{acknowledgments}

\bibliography{everything2,ers-ref}

\begin{thebibliography}{}
\expandafter\ifx\csname natexlab\endcsname\relax\def\natexlab#1{#1}\fi
\providecommand{\url}[1]{\href{#1}{#1}}
\providecommand{\dodoi}[1]{doi:~\href{http://doi.org/#1}{\nolinkurl{#1}}}
\providecommand{\doeprint}[1]{\href{http://ascl.net/#1}{\nolinkurl{http://ascl.net/#1}}}
\providecommand{\doarXiv}[1]{\href{https://arxiv.org/abs/#1}{\nolinkurl{https://arxiv.org/abs/#1}}}

\bibitem[{M.~J. {Barlow} {et~al.}(2010){Barlow}, {Krause}, {Swinyard}, {Sibthorpe}, {Besel}, {Wesson}, {Ivison}, {Dunne}, {Gear}, {Gomez}, {Hargrave}, {Henning}, {Leeks}, {Lim}, {Olofsson}, \& {Polehampton}}]{Barlow_2010}
{Barlow}, M.~J., {Krause}, O., {Swinyard}, B.~M., {et~al.} 2010, \bibinfo{title}{{A Herschel PACS and SPIRE study of the dust content of the Cassiopeia A supernova remnant},} A\&A, 518, L138, \dodoi{10.1051/0004-6361/201014585}

\bibitem[{F. {Bertoldi} {et~al.}(2003){Bertoldi}, {Carilli}, {Cox}, {Fan}, {Strauss}, {Beelen}, {Omont}, \& {Zylka}}]{Bertoldi_2003}
{Bertoldi}, F., {Carilli}, C.~L., {Cox}, P., {et~al.} 2003, \bibinfo{title}{{Dust emission from the most distant quasars},} A\&A, 406, L55, \dodoi{10.1051/0004-6361:20030710}

\bibitem[{A. {Bevan} \& M.~J. {Barlow}(2016){Bevan} \& {Barlow}}]{2016MNRAS.456.1269B}
{Bevan}, A., \& {Barlow}, M.~J. 2016, \bibinfo{title}{{Modelling supernova line profile asymmetries to determine ejecta dust masses: SN 1987A from days 714 to 3604},} \mnras, 456, 1269, \dodoi{10.1093/mnras/stv2651}

\bibitem[{H. {Bushouse} {et~al.}(2024){Bushouse}, {Eisenhamer}, {Dencheva}, {Davies}, {Greenfield}, {Morrison}, {Hodge}, {Simon}, {Grumm}, {Droettboom}, {Slavich}, {Sosey}, {Pauly}, {Miller}, {Jedrzejewski}, {Hack}, {Davis}, {Crawford}, {Law}, {Gordon}, {Regan}, {Cara}, {MacDonald}, {Bradley}, {Shanahan}, {Jamieson}, {Teodoro}, {Williams}, \& {Pena-Guerrero}}]{bushouse24}
{Bushouse}, H., {Eisenhamer}, J., {Dencheva}, N., {et~al.} 2024, \bibinfo{title}{{JWST Calibration Pipeline},}, 1.15.1 Zenodo, \dodoi{10.5281/zenodo.6984365}

\bibitem[{P. {Chandra} {et~al.}(2005){Chandra}, {Ray}, {Schlegel}, {Sutaria}, \& {Pietsch}}]{2005ApJ...629..933C}
{Chandra}, P., {Ray}, A., {Schlegel}, E.~M., {Sutaria}, F.~K., \& {Pietsch}, W. 2005, \bibinfo{title}{{Chandra's Tryst with SN 1995N},} \apj, 629, 933, \dodoi{10.1086/431573}

\bibitem[{P. {Chandra} {et~al.}(2009){Chandra}, {Stockdale}, {Chevalier}, {Van Dyk}, {Ray}, {Kelley}, {Weiler}, {Panagia}, \& {Sramek}}]{2009ApJ...690.1839C}
{Chandra}, P., {Stockdale}, C.~J., {Chevalier}, R.~A., {et~al.} 2009, \bibinfo{title}{{Eleven Years of Radio Monitoring of the type IIn Supernova SN 1995N},} \apj, 690, 1839, \dodoi{10.1088/0004-637X/690/2/1839}

\bibitem[{R.~A. {Chevalier} \& C. {Fransson}(1992){Chevalier} \& {Fransson}}]{1992ApJ...395..540C}
{Chevalier}, R.~A., \& {Fransson}, C. 1992, \bibinfo{title}{{Pulsar Nebulae in Supernovae},} \apj, 395, 540, \dodoi{10.1086/171674}

\bibitem[{C.~R. {Choban} {et~al.}(2025){Choban}, {Salim}, {Kere{\v{s}}}, {Hayward}, \& {Sandstrom}}]{2025MNRAS.tmp..117C}
{Choban}, C.~R., {Salim}, S., {Kere{\v{s}}}, D., {Hayward}, C.~C., \& {Sandstrom}, K.~M. 2025, \bibinfo{title}{{A dusty dawn: Galactic dust buildup at z {\ensuremath{\gtrsim}} 5},} \mnras, \dodoi{10.1093/mnras/staf118}

\bibitem[{N.~N. {Chugai} \& R.~A. {Chevalier}(2006){Chugai} \& {Chevalier}}]{2006ApJ...641.1051C}
{Chugai}, N.~N., \& {Chevalier}, R.~A. 2006, \bibinfo{title}{{Late Emission from the Type Ib/c SN 2001em: Overtaking the Hydrogen Envelope},} \apj, 641, 1051, \dodoi{10.1086/500539}

\bibitem[{N.~N. {Chugai} \& I.~J. {Danziger}(1994){Chugai} \& {Danziger}}]{1994MNRAS.268..173C}
{Chugai}, N.~N., \& {Danziger}, I.~J. 1994, \bibinfo{title}{{SN 1988Z: low-mass ejecta colliding with the clumpy wind?},} \mnras, 268, 173, \dodoi{10.1093/mnras/268.1.173}

\bibitem[{I. De~Looze {et~al.}(2017)De~Looze, Barlow, Swinyard, Rho, Gomez, Matsuura, \& Wesson}]{De_Looze_2017}
De~Looze, I., Barlow, M.~J., Swinyard, B.~M., {et~al.} 2017, \bibinfo{title}{The dust mass in Cassiopeia A from a spatially resolvedHerschelanalysis,} MNRAS, 465, 3309, \dodoi{10.1093/mnras/stw2837}

\bibitem[{E. {Dwek} \& R.~G. {Arendt}(2015){Dwek} \& {Arendt}}]{2015ApJ...810...75D}
{Dwek}, E., \& {Arendt}, R.~G. 2015, \bibinfo{title}{{The Evolution of Dust Mass in the Ejecta of SN1987A},} \apj, 810, 75, \dodoi{10.1088/0004-637X/810/1/75}

\bibitem[{E. {Dwek} {et~al.}(2007){Dwek}, {Galliano}, \& {Jones}}]{Dwek_2007}
{Dwek}, E., {Galliano}, F., \& {Jones}, A.~P. 2007, \bibinfo{title}{{The Evolution of Dust in the Early Universe with Applications to the Galaxy SDSS J1148+5251},} ApJ, 662, 927, \dodoi{10.1086/518430}

\bibitem[{E. {Dwek} {et~al.}(2019){Dwek}, {Sarangi}, \& {Arendt}}]{2019ApJ...871L..33D}
{Dwek}, E., {Sarangi}, A., \& {Arendt}, R.~G. 2019, \bibinfo{title}{{The Evolution of Dust Opacity in Core Collapse Supernovae and the Rapid Formation of Dust in Their Ejecta},} \apjl, 871, L33, \dodoi{10.3847/2041-8213/aaf9a8}

\bibitem[{E. Dwek {et~al.}(2014)Dwek, Staguhn, Arendt, Kovacks, Su, \& Benford}]{Dwek_2014}
Dwek, E., Staguhn, J., Arendt, R.~G., {et~al.} 2014, \bibinfo{title}{DUST FORMATION, EVOLUTION, AND OBSCURATION EFFECTS IN THE VERY HIGH-REDSHIFT UNIVERSE,} ApJ, 788, L30, \dodoi{10.1088/2041-8205/788/2/l30}

\bibitem[{B. {Ercolano} {et~al.}(2005){Ercolano}, {Barlow}, \& {Storey}}]{2005MNRAS.362.1038E}
{Ercolano}, B., {Barlow}, M.~J., \& {Storey}, P.~J. 2005, \bibinfo{title}{{The dusty MOCASSIN: fully self-consistent 3D photoionization and dust radiative transfer models},} \mnras, 362, 1038, \dodoi{10.1111/j.1365-2966.2005.09381.x}

\bibitem[{S.~M. {Faber} {et~al.}(2003){Faber}, {Phillips}, {Kibrick}, {Alcott}, {Allen}, {Burrous}, {Cantrall}, {Clarke}, {Coil}, {Cowley}, {Davis}, {Deich}, {Dietsch}, {Gilmore}, {Harper}, {Hilyard}, {Lewis}, {McVeigh}, {Newman}, {Osborne}, {Schiavon}, {Stover}, {Tucker}, {Wallace}, {Wei}, {Wirth}, \& {Wright}}]{2003SPIE.4841.1657F}
{Faber}, S.~M., {Phillips}, A.~C., {Kibrick}, R.~I., {et~al.} 2003, in Society of Photo-Optical Instrumentation Engineers (SPIE) Conference Series, Vol. 4841, Instrument Design and Performance for Optical/Infrared Ground-based Telescopes, ed. M.~{Iye} \& A.~F.~M. {Moorwood}, 1657--1669, \dodoi{10.1117/12.460346}

\bibitem[{A.~V. {Filippenko}(1997){Filippenko}}]{1997ARA&A..35..309F}
{Filippenko}, A.~V. 1997, \bibinfo{title}{{Optical Spectra of Supernovae},} \araa, 35, 309, \dodoi{10.1146/annurev.astro.35.1.309}

\bibitem[{A.~V. {Filippenko} \& T. {Matheson}(1993){Filippenko} \& {Matheson}}]{1993IAUC.5788....1F}
{Filippenko}, A.~V., \& {Matheson}, T. 1993, \bibinfo{title}{{Supernova 1993N in UGC 5695},} \iaucirc, 5788, 1

\bibitem[{A.~V. {Filippenko} \& T. {Matheson}(1994){Filippenko} \& {Matheson}}]{1994IAUC.5924....1F}
{Filippenko}, A.~V., \& {Matheson}, T. 1994, \bibinfo{title}{{Supernova 1993N in UGC 5695},} \iaucirc, 5924, 1

\bibitem[{D. {Foreman-Mackey} {et~al.}(2013){Foreman-Mackey}, {Hogg}, {Lang}, \& {Goodman}}]{2013PASP..125..306F}
{Foreman-Mackey}, D., {Hogg}, D.~W., {Lang}, D., \& {Goodman}, J. 2013, \bibinfo{title}{{emcee: The MCMC Hammer},} \pasp, 125, 306, \dodoi{10.1086/670067}

\bibitem[{D.~W. {Fox} {et~al.}(2000){Fox}, {Lewin}, {Fabian}, {Iwasawa}, {Terlevich}, {Zimmermann}, {Aschenbach}, {Weiler}, {Van Dyk}, {Chevalier}, {Rutledge}, {Inoue}, \& {Uno}}]{2000MNRAS.319.1154F}
{Fox}, D.~W., {Lewin}, W.~H.~G., {Fabian}, A., {et~al.} 2000, \bibinfo{title}{{The X-ray spectrum and light curve of Supernova 1995N},} \mnras, 319, 1154, \dodoi{10.1046/j.1365-8711.2000.03941.x}

\bibitem[{C. {Fransson} {et~al.}(2002){Fransson}, {Chevalier}, {Filippenko}, {Leibundgut}, {Barth}, {Fesen}, {Kirshner}, {Leonard}, {Li}, {Lundqvist}, {Sollerman}, \& {Van Dyk}}]{2002ApJ...572..350F}
{Fransson}, C., {Chevalier}, R.~A., {Filippenko}, A.~V., {et~al.} 2002, \bibinfo{title}{{Optical and Ultraviolet Spectroscopy of SN 1995N: Evidence for Strong Circumstellar Interaction},} \apj, 572, 350, \dodoi{10.1086/340295}

\bibitem[{C. {Gall} {et~al.}(2014){Gall}, {Hjorth}, {Watson}, {Dwek}, {Maund}, {Fox}, {Leloudas}, {Malesani}, \& {Day-Jones}}]{2014Natur.511..326G}
{Gall}, C., {Hjorth}, J., {Watson}, D., {et~al.} 2014, \bibinfo{title}{{Rapid formation of large dust grains in the luminous supernova 2010jl},} \nat, 511, 326, \dodoi{10.1038/nature13558}

\bibitem[{P. {Garnavich} {et~al.}(1995){Garnavich}, {Challis}, \& {Berlind}}]{1995IAUC.6174....1G}
{Garnavich}, P., {Challis}, P., \& {Berlind}, P. 1995, \bibinfo{title}{{Supernova 1995N in MCG -02-38-017},} \iaucirc, 6174, 1

\bibitem[{C.~L. Gerardy {et~al.}(2002)Gerardy, Fesen, Nomoto, Garnavich, Jha, Challis, Kirshner, H{\"o}flich, \& Wheeler}]{Gerardy_2002}
Gerardy, C.~L., Fesen, R.~A., Nomoto, K., {et~al.} 2002, \bibinfo{title}{Extraordinary Late-Time Infrared Emission of Type IIn Supernovae*,} The Astrophysical Journal, 575, 1007, \dodoi{10.1086/341430}

\bibitem[{C.~L. {Gerardy} {et~al.}(2002){Gerardy}, {Fesen}, {Nomoto}, {Garnavich}, {Jha}, {Challis}, {Kirshner}, {H{\"o}flich}, \& {Wheeler}}]{2002ApJ...575.1007G}
{Gerardy}, C.~L., {Fesen}, R.~A., {Nomoto}, K., {et~al.} 2002, \bibinfo{title}{{Extraordinary Late-Time Infrared Emission of Type IIn Supernovae},} \apj, 575, 1007, \dodoi{10.1086/341430}

\bibitem[{H.~L. {Gomez} {et~al.}(2012){Gomez}, {Krause}, {Barlow}, {Swinyard}, {Owen}, {Clark}, {Matsuura}, {Gomez}, {Rho}, {Besel}, {Bouwman}, {Gear}, {Henning}, {Ivison}, {Polehampton}, \& {Sibthorpe}}]{Gomez_12}
{Gomez}, H.~L., {Krause}, O., {Barlow}, M.~J., {et~al.} 2012, \bibinfo{title}{{A Cool Dust Factory in the Crab Nebula: A Herschel Study of the Filaments},} ApJ, 760, 96, \dodoi{10.1088/0004-637X/760/1/96}

\bibitem[{A.~H. {Harutyunyan} {et~al.}(2008){Harutyunyan}, {Pfahler}, {Pastorello}, {Taubenberger}, {Turatto}, {Cappellaro}, {Benetti}, {Elias-Rosa}, {Navasardyan}, {Valenti}, {Stanishev}, {Patat}, {Riello}, {Pignata}, \& {Hillebrandt}}]{2008A&A...488..383H}
{Harutyunyan}, A.~H., {Pfahler}, P., {Pastorello}, A., {et~al.} 2008, \bibinfo{title}{{ESC supernova spectroscopy of non-ESC targets},} \aap, 488, 383, \dodoi{10.1051/0004-6361:20078859}

\bibitem[{T. {Henning}(2010){Henning}}]{2010ARA&A..48...21H}
{Henning}, T. 2010, \bibinfo{title}{{Cosmic Silicates},} \araa, 48, 21, \dodoi{10.1146/annurev-astro-081309-130815}

\bibitem[{A. {Jerkstrand} {et~al.}(2012){Jerkstrand}, {Fransson}, {Maguire}, {Smartt}, {Ergon}, \& {Spyromilio}}]{2012A&A...546A..28J}
{Jerkstrand}, A., {Fransson}, C., {Maguire}, K., {et~al.} 2012, \bibinfo{title}{{The progenitor mass of the Type IIP supernova SN 2004et from late-time spectral modeling},} \aap, 546, A28, \dodoi{10.1051/0004-6361/201219528}

\bibitem[{R. {Kotak} {et~al.}(2006){Kotak}, {Meikle}, {Pozzo}, {van Dyk}, {Farrah}, {Fesen}, {Filippenko}, {Foley}, {Fransson}, {Gerardy}, {H{\"o}flich}, {Lundqvist}, {Mattila}, {Sollerman}, \& {Wheeler}}]{2006ApJ...651L.117K}
{Kotak}, R., {Meikle}, P., {Pozzo}, M., {et~al.} 2006, \bibinfo{title}{{Spitzer Measurements of Atomic and Molecular Abundances in the Type IIP SN 2005af},} \apjl, 651, L117, \dodoi{10.1086/509655}

\bibitem[{W.~H.~G. {Lewin} {et~al.}(1996){Lewin}, {Zimmermann}, \& {Aschenbach}}]{1996IAUC.6445....1L}
{Lewin}, W.~H.~G., {Zimmermann}, H.~U., \& {Aschenbach}, B. 1996, \bibinfo{title}{{Supernova 1995N in MCG -2-38-017},} \iaucirc, 6445, 1

\bibitem[{W. {Li} {et~al.}(2002){Li}, {Filippenko}, {Van Dyk}, {Hu}, {Qiu}, {Modjaz}, \& {Leonard}}]{2002PASP..114..403L}
{Li}, W., {Filippenko}, A.~V., {Van Dyk}, S.~D., {et~al.} 2002, \bibinfo{title}{{A Hubble Space Telescope Snapshot Survey of Nearby Supernovae},} \pasp, 114, 403, \dodoi{10.1086/342493}

\bibitem[{J.~S. {Mathis} {et~al.}(1977){Mathis}, {Rumpl}, \& {Nordsieck}}]{1977ApJ...217..425M}
{Mathis}, J.~S., {Rumpl}, W., \& {Nordsieck}, K.~H. 1977, \bibinfo{title}{{The size distribution of interstellar grains},} \apj, 217, 425, \dodoi{10.1086/155591}

\bibitem[{M. {Matsuura} {et~al.}(2011{\natexlab{a}}){Matsuura}, {Dwek}, {Meixner}, {Otsuka}, {Babler}, {Barlow}, {Roman-Duval}, {Engelbracht}, {Sandstrom}, {Laki{\'c}evi{\'c}}, {van Loon}, {Sonneborn}, {Clayton}, {Long}, {Lundqvist}, {Nozawa}, {Gordon}, {Hony}, {Panuzzo}, {Okumura}, {Misselt}, {Montiel}, \& {Sauvage}}]{Matsuura_11}
{Matsuura}, M., {Dwek}, E., {Meixner}, M., {et~al.} 2011{\natexlab{a}}, \bibinfo{title}{{Herschel Detects a Massive Dust Reservoir in Supernova 1987A},} Science, 333, 1258, \dodoi{10.1126/science.1205983}

\bibitem[{M. {Matsuura} {et~al.}(2011{\natexlab{b}}){Matsuura}, {Dwek}, {Meixner}, {Otsuka}, {Babler}, {Barlow}, {Roman-Duval}, {Engelbracht}, {Sandstrom}, {Laki{\'c}evi{\'c}}, {van Loon}, {Sonneborn}, {Clayton}, {Long}, {Lundqvist}, {Nozawa}, {Gordon}, {Hony}, {Panuzzo}, {Okumura}, {Misselt}, {Montiel}, \& {Sauvage}}]{2011Sci...333.1258M}
{Matsuura}, M., {Dwek}, E., {Meixner}, M., {et~al.} 2011{\natexlab{b}}, \bibinfo{title}{{Herschel Detects a Massive Dust Reservoir in Supernova 1987A},} Science, 333, 1258, \dodoi{10.1126/science.1205983}

\bibitem[{M. {Matsuura} {et~al.}(2015){Matsuura}, {Dwek}, {Barlow}, {Babler}, {Baes}, {Meixner}, {Cernicharo}, {Clayton}, {Dunne}, {Fransson}, {Fritz}, {Gear}, {Gomez}, {Groenewegen}, {Indebetouw}, {Ivison}, {Jerkstrand}, {Lebouteiller}, {Lim}, {Lundqvist}, {Pearson}, {Roman-Duval}, {Royer}, {Staveley-Smith}, {Swinyard}, {van Hoof}, {van Loon}, {Verstappen}, {Wesson}, {Zanardo}, {Blommaert}, {Decin}, {Reach}, {Sonneborn}, {Van de Steene}, \& {Yates}}]{Matsuura_15}
{Matsuura}, M., {Dwek}, E., {Barlow}, M.~J., {et~al.} 2015, \bibinfo{title}{{A Stubbornly Large Mass of Cold Dust in the Ejecta of Supernova 1987A},} ApJ, 800, 50, \dodoi{10.1088/0004-637X/800/1/50}

\bibitem[{W.~P.~S. {Meikle} {et~al.}(2007){Meikle}, {Mattila}, {Pastorello}, {Gerardy}, {Kotak}, {Sollerman}, {Van Dyk}, {Farrah}, {Filippenko}, {H{\"o}flich}, {Lundqvist}, {Pozzo}, \& {Wheeler}}]{2007ApJ...665..608M}
{Meikle}, W.~P.~S., {Mattila}, S., {Pastorello}, A., {et~al.} 2007, \bibinfo{title}{{A Spitzer Space Telescope Study of SN 2003gd: Still No Direct Evidence that Core-Collapse Supernovae are Major Dust Factories},} \apj, 665, 608, \dodoi{10.1086/519733}

\bibitem[{D. {Milisavljevic} \& R.~A. {Fesen}(2008){Milisavljevic} \& {Fesen}}]{2008ApJ...677..306M}
{Milisavljevic}, D., \& {Fesen}, R.~A. 2008, \bibinfo{title}{{The Nature of the Ultraluminous Oxygen-Rich Supernova Remnant in NGC 4449},} \apj, 677, 306, \dodoi{10.1086/528929}

\bibitem[{D. {Milisavljevic} \& R.~A. {Fesen}(2017){Milisavljevic} \& {Fesen}}]{2017hsn..book.2211M}
{Milisavljevic}, D., \& {Fesen}, R.~A. 2017, in Handbook of Supernovae, ed. A.~W. {Alsabti} \& P.~{Murdin}, 2211, \dodoi{10.1007/978-3-319-21846-5_97}

\bibitem[{D. {Milisavljevic} {et~al.}(2015){Milisavljevic}, {Margutti}, {Kamble}, {Patnaude}, {Raymond}, {Eldridge}, {Fong}, {Bietenholz}, {Challis}, {Chornock}, {Drout}, {Fransson}, {Fesen}, {Grindlay}, {Kirshner}, {Lunnan}, {Mackey}, {Miller}, {Parrent}, {Sanders}, {Soderberg}, \& {Zauderer}}]{2015ApJ...815..120M}
{Milisavljevic}, D., {Margutti}, R., {Kamble}, A., {et~al.} 2015, \bibinfo{title}{{Metamorphosis of SN 2014C: Delayed Interaction between a Hydrogen Poor Core-collapse Supernova and a Nearby Circumstellar Shell},} \apj, 815, 120, \dodoi{10.1088/0004-637X/815/2/120}

\bibitem[{J.~S. {Miller} \& R.~P.~S. {Stone}(1993){Miller} \& {Stone}}]{miller93}
{Miller}, J.~S., \& {Stone}, R.~P.~S. 1993, Tech. Rep.~66

\bibitem[{J. {Mueller} {et~al.}(1993{\natexlab{a}}){Mueller}, {Brewer}, {Cappellaro}, \& {della Valle}}]{1993IAUC.5784....1M}
{Mueller}, J., {Brewer}, C., {Cappellaro}, E., \& {della Valle}, M. 1993{\natexlab{a}}, \bibinfo{title}{{Supernova 1993N in UGC 5695},} \iaucirc, 5784, 1

\bibitem[{J. {Mueller} {et~al.}(1993{\natexlab{b}}){Mueller}, {Mould}, \& {Smail}}]{1993IAUC.5791....3M}
{Mueller}, J., {Mould}, J.~R., \& {Smail}, I. 1993{\natexlab{b}}, \bibinfo{title}{{Supernova 1993N in UGC 5695},} \iaucirc, 5791, 3

\bibitem[{J.~B. {Oke} {et~al.}(1995){Oke}, {Cohen}, {Carr}, {Cromer}, {Dingizian}, {Harris}, {Labrecque}, {Lucinio}, {Schaal}, {Epps}, \& {Miller}}]{1995PASP..107..375O}
{Oke}, J.~B., {Cohen}, J.~G., {Carr}, M., {et~al.} 1995, \bibinfo{title}{{The Keck Low-Resolution Imaging Spectrometer},} \pasp, 107, 375, \dodoi{10.1086/133562}

\bibitem[{P.~J. {Owen} \& M.~J. {Barlow}(2015){Owen} \& {Barlow}}]{Owen_15}
{Owen}, P.~J., \& {Barlow}, M.~J. 2015, \bibinfo{title}{{The Dust and Gas Content of the Crab Nebula},} ApJ, 801, 141, \dodoi{10.1088/0004-637X/801/2/141}

\bibitem[{N. {Panagia} {et~al.}(2000){Panagia}, {Weiler}, {Lacey}, {Montes}, {Sramek}, \& {van Dyk}}]{2000MmSAI..71..331P}
{Panagia}, N., {Weiler}, K.~W., {Lacey}, C., {et~al.} 2000, \bibinfo{title}{{Radio studies of supernovae.},} \memsai, 71, 331

\bibitem[{A. {Pastorello} {et~al.}(2005){Pastorello}, {Aretxaga}, {Zampieri}, {Mucciarelli}, \& {Benetti}}]{2005ASPC..342..285P}
{Pastorello}, A., {Aretxaga}, I., {Zampieri}, L., {Mucciarelli}, P., \& {Benetti}, S. 2005, in Astronomical Society of the Pacific Conference Series, Vol. 342, 1604-2004: Supernovae as Cosmological Lighthouses, ed. M.~{Turatto}, S.~{Benetti}, L.~{Zampieri}, \& W.~{Shea}, 285, \dodoi{10.48550/arXiv.astro-ph/0504116}

\bibitem[{A. {Pastorello} {et~al.}(2011){Pastorello}, {Benetti}, {Bufano}, {Kankare}, {Mattila}, {Turatto}, \& {Cupani}}]{2011AN....332..266P}
{Pastorello}, A., {Benetti}, S., {Bufano}, F., {et~al.} 2011, \bibinfo{title}{{Supernovae interacting with a circumstellar medium: New observations with X-shooter},} Astronomische Nachrichten, 332, 266, \dodoi{10.1002/asna.201111532}

\bibitem[{D. {Patnaude} {et~al.}(2024){Patnaude}, {Weil}, {Fesen}, {Milisavljevic}, \& {Kraft}}]{2024arXiv241213024P}
{Patnaude}, D., {Weil}, K., {Fesen}, R., {Milisavljevic}, D., \& {Kraft}, R. 2024, \bibinfo{title}{{Late-Time Optical and X-ray Emission Evolution of the Oxygen-Rich SN 1996cr},} arXiv e-prints, arXiv:2412.13024, \dodoi{10.48550/arXiv.2412.13024}

\bibitem[{D.~A. {Perley}(2019){Perley}}]{2019PASP..131h4503P}
{Perley}, D.~A. 2019, \bibinfo{title}{{Fully Automated Reduction of Longslit Spectroscopy with the Low Resolution Imaging Spectrometer at the Keck Observatory},} \pasp, 131, 084503, \dodoi{10.1088/1538-3873/ab215d}

\bibitem[{C. {Pollas} {et~al.}(1995){Pollas}, {Albanese}, {Benetti}, {Bouchet}, \& {Schwarz}}]{1995IAUC.6170....1P}
{Pollas}, C., {Albanese}, D., {Benetti}, S., {Bouchet}, P., \& {Schwarz}, H. 1995, \bibinfo{title}{{Supernova 1995N in MCG -02-38-017},} \iaucirc, 6170, 1

\bibitem[{B.~E. {Schaefer}(2001){Schaefer}}]{2001IAUC.7626....3S}
{Schaefer}, B.~E. 2001, \bibinfo{title}{{Supernova 1995N in MCG -02-38-017},} \iaucirc, 7626, 3

\bibitem[{B.~E. {Schaefer} \& B. {Roscherr}(1999){Schaefer} \& {Roscherr}}]{1999IAUC.7141....3S}
{Schaefer}, B.~E., \& {Roscherr}, B. 1999, \bibinfo{title}{{Supernovae 1995N and 1997ab},} \iaucirc, 7141, 3

\bibitem[{M. {Shahbandeh} {et~al.}(2023){Shahbandeh}, {Sarangi}, {Temim}, {Szalai}, {Fox}, {Tinyanont}, {Dwek}, {Dessart}, {Filippenko}, {Brink}, {Foley}, {Jencson}, {Pierel}, {Zs{\'\i}ros}, {Rest}, {Zheng}, {Andrews}, {Clayton}, {De}, {Engesser}, {Gezari}, {Gomez}, {Gonzaga}, {Johansson}, {Kasliwal}, {Lau}, {De Looze}, {Marston}, {Milisavljevic}, {O'Steen}, {Siebert}, {Skrutskie}, {Smith}, {Strolger}, {Van Dyk}, {Wang}, {Williams}, {Williams}, {Xiao}, \& {Yang}}]{shahbandeh23}
{Shahbandeh}, M., {Sarangi}, A., {Temim}, T., {et~al.} 2023, \bibinfo{title}{{JWST observations of dust reservoirs in type IIP supernovae 2004et and 2017eaw},} \mnras, 523, 6048, \dodoi{10.1093/mnras/stad1681}

\bibitem[{M. {Shahbandeh} {et~al.}(2024){Shahbandeh}, {Fox}, {Temim}, {Dwek}, {Sarangi}, {Smith}, {Dessart}, {Nickson}, {Engesser}, {Filippenko}, {Brink}, {Zheng}, {Szalai}, {Johansson}, {Rest}, {Van Dyk}, {Andrews}, {Ashall}, {Clayton}, {De Looze}, {Derkacy}, {Dulude}, {Foley}, {Gezari}, {Gomez}, {Gonzaga}, {Indukuri}, {Jencson}, {Kasliwal}, {Lane}, {Lau}, {Law}, {Marston}, {Milisavljevic}, {O'Steen}, {Pierel}, {Siebert}, {Skrutskie}, {Strolger}, {Tinyanont}, {Wang}, {Williams}, {Xiao}, {Yang}, \& {Zs{\'\i}ros}}]{2024arXiv241009142S}
{Shahbandeh}, M., {Fox}, O.~D., {Temim}, T., {et~al.} 2024, \bibinfo{title}{{JWST/MIRI Observations of Newly Formed Dust in the Cold, Dense Shell of the Type IIn SN 2005ip},} arXiv e-prints, arXiv:2410.09142, \dodoi{10.48550/arXiv.2410.09142}

\bibitem[{J.~M. {Silverman} {et~al.}(2012){Silverman}, {Foley}, {Filippenko}, {Ganeshalingam}, {Barth}, {Chornock}, {Griffith}, {Kong}, {Lee}, {Leonard}, {Matheson}, {Miller}, {Steele}, {Barris}, {Bloom}, {Cobb}, {Coil}, {Desroches}, {Gates}, {Ho}, {Jha}, {Kandrashoff}, {Li}, {Mandel}, {Modjaz}, {Moore}, {Mostardi}, {Papenkova}, {Park}, {Perley}, {Poznanski}, {Reuter}, {Scala}, {Serduke}, {Shields}, {Swift}, {Tonry}, {Van Dyk}, {Wang}, \& {Wong}}]{2012MNRAS.425.1789S}
{Silverman}, J.~M., {Foley}, R.~J., {Filippenko}, A.~V., {et~al.} 2012, \bibinfo{title}{{Berkeley Supernova Ia Program - I. Observations, data reduction and spectroscopic sample of 582 low-redshift Type Ia supernovae},} \mnras, 425, 1789, \dodoi{10.1111/j.1365-2966.2012.21270.x}

\bibitem[{R.~A. {Stathakis} \& E.~M. {Sadler}(1991){Stathakis} \& {Sadler}}]{1991MNRAS.250..786S}
{Stathakis}, R.~A., \& {Sadler}, E.~M. 1991, \bibinfo{title}{{What was supernova 1988Z?},} \mnras, 250, 786, \dodoi{10.1093/mnras/250.4.786}

\bibitem[{T. {Szalai} {et~al.}(2011){Szalai}, {Vink{\'o}}, {Balog}, {G{\'a}sp{\'a}r}, {Block}, \& {Kiss}}]{2011A&A...527A..61S}
{Szalai}, T., {Vink{\'o}}, J., {Balog}, Z., {et~al.} 2011, \bibinfo{title}{{Dust formation in the ejecta of the type II-P supernova 2004dj},} \aap, 527, A61, \dodoi{10.1051/0004-6361/201015624}

\bibitem[{T. {Szalai} {et~al.}(2021){Szalai}, {Fox}, {Arendt}, {Dwek}, {Andrews}, {Clayton}, {Filippenko}, {Johansson}, {Kelly}, {Krafton}, {Marston}, {Mauerhan}, \& {Van Dyk}}]{2021ApJ...919...17S}
{Szalai}, T., {Fox}, O.~D., {Arendt}, R.~G., {et~al.} 2021, \bibinfo{title}{{Spitzer's Last Look at Extragalactic Explosions: Long-term Evolution of Interacting Supernovae},} \apj, 919, 17, \dodoi{10.3847/1538-4357/ac0e2b}

\bibitem[{T. Temim \& E. Dwek(2013)Temim \& Dwek}]{Temim_2013}
Temim, T., \& Dwek, E. 2013, \bibinfo{title}{THE IMPORTANCE OF PHYSICAL MODELS FOR DERIVING DUST MASSES AND GRAIN SIZE DISTRIBUTIONS IN SUPERNOVA EJECTA. I. RADIATIVELY HEATED DUST IN THE CRAB NEBULA,} ApJ, 774, 8, \dodoi{10.1088/0004-637x/774/1/8}

\bibitem[{T. Temim {et~al.}(2017)Temim, Dwek, Arendt, Borkowski, Reynolds, Slane, Gelfand, \& Raymond}]{Temim_2017}
Temim, T., Dwek, E., Arendt, R.~G., {et~al.} 2017, \bibinfo{title}{A Massive Shell of Supernova-formed Dust in SNR G54.1+0.3,} ApJ, 836, 129, \dodoi{10.3847/1538-4357/836/1/129}

\bibitem[{M. {Turatto} {et~al.}(1993){Turatto}, {Cappellaro}, {Danziger}, {Benetti}, {Gouiffes}, \& {della Valle}}]{1993MNRAS.262..128T}
{Turatto}, M., {Cappellaro}, E., {Danziger}, I.~J., {et~al.} 1993, \bibinfo{title}{{The type II supernova 1988Z in MCG +03-28-022 : increasingevidence of interaction of supernova ejecta with a circumstellar wind.},} \mnras, 262, 128, \dodoi{10.1093/mnras/262.1.128}

\bibitem[{S.~D. {Van Dyk}(2013){Van Dyk}}]{2013AJ....145..118V}
{Van Dyk}, S.~D. 2013, \bibinfo{title}{{Late-time Dust Emission from the Type IIn Supernova 1995N},} \aj, 145, 118, \dodoi{10.1088/0004-6256/145/5/118}

\bibitem[{S.~D. {Van Dyk} {et~al.}(1996){Van Dyk}, {Sramek}, {Weiler}, {Montes}, \& {Panagia}}]{1996IAUC.6386....1V}
{Van Dyk}, S.~D., {Sramek}, R.~A., {Weiler}, K.~W., {Montes}, M.~J., \& {Panagia}, N. 1996, \bibinfo{title}{{Supernova 1995N in MCG -2-38-017},} \iaucirc, 6386, 1

\bibitem[{R. {Wesson} {et~al.}(2015){Wesson}, {Barlow}, {Matsuura}, \& {Ercolano}}]{2015MNRAS.446.2089W}
{Wesson}, R., {Barlow}, M.~J., {Matsuura}, M., \& {Ercolano}, B. 2015, \bibinfo{title}{{The timing and location of dust formation in the remnant of SN 1987A},} \mnras, 446, 2089, \dodoi{10.1093/mnras/stu2250}

\bibitem[{R. {Wesson} {et~al.}(2023){Wesson}, {Bevan}, {Barlow}, {De Looze}, {Matsuura}, {Clayton}, \& {Andrews}}]{2023MNRAS.525.4928W}
{Wesson}, R., {Bevan}, A.~M., {Barlow}, M.~J., {et~al.} 2023, \bibinfo{title}{{Evidence for late-time dust formation in the ejecta of supernova SN 1995N from emission-line asymmetries},} \mnras, 525, 4928, \dodoi{10.1093/mnras/stad2505}

\bibitem[{D.~H. {Wooden} {et~al.}(1993){Wooden}, {Rank}, {Bregman}, {Witteborn}, {Tielens}, {Cohen}, {Pinto}, \& {Axelrod}}]{1993ApJS...88..477W}
{Wooden}, D.~H., {Rank}, D.~M., {Bregman}, J.~D., {et~al.} 1993, \bibinfo{title}{{Airborne spectrophotometry of SN 1987A from 1.7 to 12.6 microns - Time history of the dust continuum and line emission},} \apjs, 88, 477, \dodoi{10.1086/191830}

\bibitem[{L. {Zampieri} {et~al.}(2005){Zampieri}, {Mucciarelli}, {Pastorello}, {Turatto}, {Cappellaro}, \& {Benetti}}]{2005MNRAS.364.1419Z}
{Zampieri}, L., {Mucciarelli}, P., {Pastorello}, A., {et~al.} 2005, \bibinfo{title}{{Simultaneous XMM-Newton and ESO VLT observations of supernova 1995N: probing the wind-ejecta interaction},} \mnras, 364, 1419, \dodoi{10.1111/j.1365-2966.2005.09671.x}

\bibitem[{S. {Zs{\'\i}ros} {et~al.}(2024){Zs{\'\i}ros}, {Szalai}, {De Looze}, {Sarangi}, {Shahbandeh}, {Fox}, {Temim}, {Milisavljevic}, {Van Dyk}, {Smith}, {Filippenko}, {Brink}, {Zheng}, {Dessart}, {Jencson}, {Johansson}, {Pierel}, {Rest}, {Tinyanont}, {Niculescu-Duvaz}, {Barlow}, {Wesson}, {Andrews}, {Clayton}, {De}, {Dwek}, {Engesser}, {Foley}, {Gezari}, {Gomez}, {Gonzaga}, {Kasliwal}, {Lau}, {Marston}, {O'Steen}, {Siebert}, {Skrutskie}, {Strolger}, {Wang}, {Williams}, {Williams}, \& {Xiao}}]{zsiros24}
{Zs{\'\i}ros}, S., {Szalai}, T., {De Looze}, I., {et~al.} 2024, \bibinfo{title}{{Serendipitous detection of the dusty Type IIL SN 1980K with JWST/MIRI},} \mnras, 529, 155, \dodoi{10.1093/mnras/stae507}

\end{thebibliography}
\end{document}